\documentclass[particles,article,accept,pdftex,moreauthors]{Definitions/mdpi}

\def\draftnote#1{{\it #1}}

\newcommand*{\citenamefont}[1]{#1}
\newcommand*{\bibnamefont}[1]{#1}
\newcommand*{\bibfnamefont}[1]{#1}
\providecommand \doibase [0]{http://dx.doi.org/}

\firstpage{144}
\pubvolume{6}
\issuenum{1}
\articlenumber{8}
\pubyear{2023}
\copyrightyear{2023}
\datereceived{8 October 2022}
\dateaccepted{19 January 2023}
\datepublished{28 January 2023} 
\hreflink{https://doi.org/10.3390/particles6010008} 

\Title{Beyond the standard model with six-dimensional spinors}

\TitleCitation{Beyond the standard model with six-dimensional spinors}

\Author{David Chester $^{1,}$*\orcidA{}, Alessio Marrani $^{2}$ and Michael Rios $^{3}$}

\AuthorNames{David Chester, Alessio Marrani and Michael Rios}

\AuthorCitation{Chester, D.; Marrani, A.; Rios, M.}

\address{%
$^{1}$ \quad Quantum Gravity Research, Topanga, CA, USA; davidc@quantumgravityresearch.org \\
$^{2}$ \quad Instituto de F\'{i}sica Teorica, Universidad de Murcia, Campus de Espinardo, E-30100, Spain; alessio.marrani@um.es \\
$^{3}$ \quad Dyonica ICMQG, Los Angeles, CA, USA; Quantum Gravity Research, Topanga, CA, USA; mike@quantumgravityresearch.org}

\corres{Correspondence: davidc@quantumgravityresearch.org}

\abstract{6D spinors with $Spin(3,3)$ symmetry are utilized to efficiently encode three generations of matter. 
$E_{8(-24)}$ is shown to contain physically relevant subgroups with representations for GUT groups, spacetime symmetries, three generations of the standard model fermions, and Higgs bosons. Pati-Salam, $SU(5)$, and $Spin(10)$ grand unified theories are found when a single generation is isolated.
For spacetime symmetries, $Spin(4,2)$ may be used for conformal symmetry, $AdS_5\rightarrow dS_4$, or simply broken to $Spin(3,1)$ of Minkowski space. Another class of representations finds $Spin(2,2)$ and can give $AdS_3$ with various GUTs. An action for three generations of fermions in the Majorana-Weyl spinor ${\bf 128}$ of $Spin(4,12)$ is found with $Spin(3)$ flavor symmetry inside $E_{8(-24)}$. The ${\bf 128}$ of $Spin(4,12)$ can be regarded as the tangent space to a particular pseudo-Riemannian form of the octo-octonionic Rosenfeld projective plane $E_{8(-24)}/Spin(4,12)= (\mathbb{O}_s\times\mathbb{O})\mathbb{P}^2$. 
}


\begin{document}


\section{Introduction}

Besides the Pati-Salam $\mathfrak{su}_4\oplus\mathfrak{su}_2\oplus\mathfrak{su}_2$ model~\cite{Pati:1974yy,Pati:2003qia,Hartmann:2014fya,Saad:2017pqj,Molinaro:2018kjz,Li:2019nvi},
the $\mathfrak{e}$-series of Lie algebras ($\mathfrak{e}_4 \sim \mathfrak{su}_5$~\cite{Georgi:1974sy,Dimopulos1981,Ibanez1982,BUCHMULLER1986,Giveon1991,Arnowitt1992,Arnowitt1993,Altarelli2001}, $\mathfrak{e}_5 \sim \mathfrak{so}_{10}$~\cite{Georgi-1975,Fritzsch:1974nn,DelAguila1981,Aulakh1983,Babu1993,Frohlich1994,Hall1994,Barr1997,
Lykken:1997eh,Plumacher:1998ex,Blazek2002,Ozer:2005dwq,King2006,Bertolini:2009qj,Feruglio2015,Babu:2016bmy,deAnda2017}, $\mathfrak{e}_6$~\cite{Gursey1980,BARBIERI1980,Mohapatra:1985xm,Hall:2001xr,Ito:2010df,Chiang:2011sj,Benli:2017eld,Schwichtenberg:2018aqc}, $\mathfrak{e}_7$~\cite{Gursey1976,LEDNICKY1979302,Kugo:1983ai}, and $\mathfrak{e}_8$~\cite{Bars1980,Baaklini:1980uq,Konshtein:1980km,Koca:1981xd,Ong:1984ej,Itoh:1985jz,Buchmuller:1985rc,Barr:1987pu,Mahapatra:1988gc,Ellwanger:1990ss,Adler:2002yg,Adler:2004uj,Lisi:2007gv,Pavsic:2008sz,Castro:2009zzd,Lisi:2010uw,Castro:2014wna,2014Douglas,Lisi:2015oja}) is used to describe various grand unification theories (GUTs). Additionally, string theory proposes an $E_8\times E_8$ heterotic theory~\cite{Green:1984sg,CANDELAS1985,Dixon:1985jw,Dixon:1986jc,Greene:1986qva}. Beyond the standard model (BSM) physics is also studied with the infinite-dimensional exceptional Lie algebras, such as $\mathfrak{e}_{10}$~\cite{PhysRevLett.89.221601,Kleinschmidt:2006dy,Kleinschmidt:2006ks,2006Houart,Palmkvist:2009bw} and $\mathfrak{e}_{11}$~\cite{West_2001,deBuyl:2006gp,Bossard:2019ksx}. This work investigates the role of $\mathfrak{e}_{8(-24)}$ as the noncompact (quaternionic) real form of $\mathfrak{e}_8$ in unification physics, not as a single GUT model, but as a single algebra that breaks to representations of GUT gauge algebras, spacetime algebras, fermions, and Higgs sectors.

Three generations of matter from 6D spinors via $\mathfrak{so}_{3,3}$ is a feature of all of the models discussed within this work. 6D physics has been successful for additional mass terms and unitarity methods~\cite{Cheung:2009dc,Bern:2010qa,Chester:2016ojq}, standard model (SM) physics~\cite{Babu:2002ti,Kojima:2017qbt}, whereas three-time physics has been discussed in supersymmetric models with multiple superparticles~\cite{Sezgin:1997gr,Bars:1997ug}, and various graviweak/graviGUT proposals~\cite{Nesti:2007ka,Alexander:2007mt,Percacci:2008zf,Das:2013xha,Froggatt:2013lba,Das:2014xga,Laperashvili:2014xea,Laperashvili:2015pea,Das:2015usa,Sidharth:2018zev} have used $\mathfrak{so}_{3,11}$ for the SM~\cite{Nesti:2009kk,Percacci:2009ij,Lisi:2010td,Lisi:2015oja}. 
In previous work, the authors have explored branes given by exceptional periodicity (EP)~\cite{Rios:2018lhc}, and all of the so-called EP algebras $\mathfrak{e}_{8(-24)}^{(n)}$ ($n\in\mathbb{Z}^{+}$) allow reductions along 3-branes or 4-branes with dual magnetic brane cohomologies encoding spinors.  The $n=3$ case in $D=27+3$ was used to propose a worldvolume interpretation for M-theory~\cite{Rios:2019rfc}.  Here, we build on the result for $n=1$ that $\mathfrak{e}_{8(-24)}$ allows for a 3-brane (with (3,3) worldvolume) to be found by breaking to $\mathfrak{so}_{4,12}\rightarrow\mathfrak{so}_{3,11}$.

The double copy finds a relationship between Yang-Mills and gravity~\cite{Bern:2008qj,Bern:2010yg,Bern:2012uf,Bern:2013yya,Bern:2015ooa,Monteiro:2014cda,Ridgway:2015fdl,ChesterEYM}. Given the recent developments on the double copy and heterotic theories~\cite{Chiodaroli:2017ehv,Azevedo:2018dgo,Cho:2019ype}, this work is complementary with these developments and the graviGUT models by finding an internal charge space and external spacetime. For $\mathfrak{e}_{8(-24)}$, breaking $\mathfrak{so}_{12,4}\rightarrow \mathfrak{so}_{3,3}\oplus \mathfrak{so}_{9,1}$ allows for the identification of 6D spacetime and 10D charge space\footnote{This internal charge space can have the same signature of the critical spacetime dimension for superstring theory. Here, we find that $\mathfrak{so}_{9,1}$ is relevant as a charge space for the SM that is in a sense dual to the $\mathfrak{so}_{10}$ GUT algebra in $\mathfrak{e}_{8(-24)}$.}. Removing a lightcone gives $\mathfrak{so}_{11,3}$ and relates to branes found in three-time supersymmetry models~\cite{Sezgin:1997gr,Bars:1997ug,Rudychev:1997ue,Rudychev:1997ui}. The benefit of three times allows for three superparticles, which can be interpreted to yield three generations of fermions.

Given the difficulties of finding UV-finite quantum gravity and its small coupling constant, GUTs were proposed to unify all of the fundamental forces besides gravity. Recently, it has been suggested that torsion allows for UV-complete fermions~\cite{Poplawski:2009su,Poplawski:2017yeo,Poplawski:2018jik}, while the Gauss-Bonnet term allows for two-loop graviton scattering~\cite{Bern:2017puu}, both of which are related to the MacDowell-Mansouri formalism~\cite{PhysRevLett.38.739,Wise:2006sm,Bjorken2013,Aydemir:2017hyf} studied in various graviGUT models~\cite{Lisi:2015oja,Krasnov:2017epi}. Grand unified theory has been studied in a supersymmetry or supergravity context~\cite{Aulakh1983,Hall1994,Plumacher:1998ex,Feruglio2015,Ito:2010df,Konshtein:1980km,Ong:1984ej,Itoh:1985jz,Ellwanger:1990ss,Adler:2004uj,Li:2019nvi}.

Using $\mathfrak{so}_{3,3}$ spinor representations allow for $\mathfrak{e}_{8(-24)}$ to describe three generations of matter with only 128 degrees of freedom (dofs) instead of 192 dofs typically used to describe three generations of SM fermions), which corrects aspects of the $\mathfrak{sl}_{2,\mathbb{C}}$ model~\cite{Lisi:2015oja}. Complaints with $E_8$ for unified theory~\cite{Distler:2009jt} do not apply to 6D spacetime, as only 128 fermions are needed for three generations~\cite{Maalampi:1988va}. While $E_8$ does not contain complex representations, the algebra can be broken to smaller algebras with complex representations, such as $\mathfrak{so}_{10}$. Real Majorana spinors exhibiting Majorana-Weyl chiral spinor decompositions with three independent complex subspaces with respect to $D=3+1$ can be found with $Cl(3,3)$, $Cl(4,4)$, $Cl(11,3)$, and $Cl(12,4)$. Additionally, the ${\bf128}$ spinor inside $\mathfrak{e}_{8(-24)}$ relates to $(\mathbb{O}_s\otimes \mathbb{O})\mathbb{P}^2$, which is not complex, but $\mathbb{O}_s$ contains three complex subalgebras for three generations of chiral spinors.

While typical GUTs study an algebra and add representations, the representation theory discussed here breaks $\mathfrak{e}_{8(-24)}$ alone into GUT algebras, spinors, spacetime algebras, and Higgs representations. $E_8$ gauge theory need not be used as a GUT in the conventional sense in this manner; nevertheless, $E_8$ can be used to encapsulate GUTs with spacetime by gauging subgroups. \\

The manuscript is ordered as follows. Next, representation theory of $SO(10)$, $SU(5)$, Pati-Salam, and $E_6$ GUTs is introduced. Section~\eqref{3massGen} explains how $E_6$ GUT fits into $E_7$ for one generation and how the magic star projection of $E_8$ (also called the $\mathfrak{g}_2$ decomposition by Mukai~\cite{Mukai}) motivates three generations~\cite{Truini:2011np,Marrani:2015sta}. It also introduces $\mathfrak{so}_{3,3}$ spacetime via a toy model from $\mathfrak{f}_{4(4)}$ that contains various spinors.  Section~\eqref{4spatialDim} discusses extensions of previous (and new) graviGUT models by breaking $\mathfrak{so}_{4,12}$ (with four spacelike dimensions) to the SM spectrum. Section~\eqref{4temporalDim} explores additional models from $\mathfrak{so}_{12,4}$ with four timelike dimensions. Concluding remarks are given in Section~\eqref{conc}. 

\subsection{A review of various grand unification models}


Gauge theories are QFTs with local symmetry whose fields are representations of the gauge group, which can include spacetime symmetries for gauge theories of gravity. The standard model and all GUT models are devoid of gravity, implying that fields may be representations of the gauge group plus spacetime symmetries. Gauge theories may be spontaneously broken when a vacuum expectation value of one of the (Higgs) fields is taken, which finds a low-energy theory whose gauge symmetry is a subalgebra. Since representation theory uniquely determines the field content of a theory \cite{Slansky}, we primarily explore representation theory of $\mathfrak{e}_{8(-24)}$ and its subalgebras throughout in order to describe a landscape of possible models. Our notation is that a semi-simple Lie algebra $\mathfrak{g}_1\oplus\mathfrak{g}_2\oplus\dots\oplus \mathfrak{a}_1$ as a direct sum of non-Abelian Lie algebras $\mathfrak{g}_i$ and Abelian algebras $\mathfrak{a}_j$ has non-Abelian representations in bold and Abelian $\mathfrak{u}_1$ charges or $\mathfrak{so}_{1,1}$ weights as subscripts, such that $({\bf a},{\bf b},\dots)_c$ corresponds to a field in the ${\bf a}$ representation of $\mathfrak{g}_1$, the ${\bf b}$ representation of $\mathfrak{g}_2$, and charge $c$ with respect to $\mathfrak{a}_1$. 

The $\mathfrak{su}_3\oplus\mathfrak{su}_2 \oplus \mathfrak{u}_1$ gauge algebra of the SM can be found from the symmetry breaking of the Pati-Salam GUT with $\mathfrak{su}_4\oplus \mathfrak{su}_2\oplus\mathfrak{su}_2$ gauge symmetry~\cite{Pati:1974yy} and the Georgi-Glashow GUT with $\mathfrak{su}_5$~\cite{Georgi:1974sy}.
Pati-Salam GUT allows for a fermionic unification of the quarks and the leptons into $({\bf4},{\bf2},{\bf1})$ and $(\overline{{\bf4}},{\bf1},{\bf2})$ representations by treating the leptons as a fourth color. Alternatively, $\mathfrak{su}_5$ unifies the bosons into a single gauge group. The fermions are placed in the $\overline{\mathbf{5}}$ and $\mathbf{10}$ representations.

Both of these GUT algebras can be unified into $\mathfrak{so}_{10}$ with ${\bf 16}\oplus\overline{{\bf 16}}$ chiral spinors for fermions, since
\begin{eqnarray}
\label{so10comm}
\mathfrak{so}_{10} &\rightarrow& \mathfrak{su}_{4,c}\oplus \mathfrak{su}_{2,L} \oplus \mathfrak{su}_{2,R} \nonumber \\
\downarrow\,\, && \qquad \quad \,\, \downarrow \\
\mathfrak{su}_5 &\rightarrow& \mathfrak{su}_{3,c}\oplus\mathfrak{su}_{2,L}\oplus\mathfrak{u}_{1,Y} \rightarrow \mathfrak{su}_{3,c}\oplus\mathfrak{u}_{1,e}, \nonumber
\end{eqnarray}
where $\mathfrak{su}_{3.c} \oplus \mathfrak{su}_{2,L} \oplus \mathfrak{u}_{1,Y}$ is the algebra associated with the SM and $\mathfrak{u}_{1,e}$ describes electromagnetism. The commutative diagram in Eq.~\eqref{so10comm}~\cite{Baez:2009dj} denotes that the same SM gauge group is found from $\mathfrak{so}_{10}$ via $\mathfrak{su}_5\oplus\mathfrak{u}_1$ or $\mathfrak{su}_4\oplus \mathfrak{su}_2\oplus \mathfrak{su}_2$. Finding $\mathfrak{u}_{1,Y}$ in $\mathfrak{su}_5$ or $\mathfrak{su}_5\oplus\mathfrak{u}_1$ differentiates between $SU(5)$ or flipped $SU(5)$ GUT~\cite{BARR1982,DERENDINGER1984,ANTONIADIS1987,TAMVAKIS1988,Antoniadis:1989zy,Lopez:1991eca,Huang:2006nu,Chen2011,Ellis:2018moe}.
For $\mathfrak{so}_{10}\rightarrow\mathfrak{su}_5\oplus\mathfrak{u}_{1}$, a right-handed neutrino is required, since the ${\bf16}$ contains ${\bf1}_{-5}$,
\begin{eqnarray}
\mathfrak{so}_{10} &\rightarrow& \mathfrak{su}_5\oplus \mathfrak{u}_{1,X} \rightarrow \mathfrak{su}_{3,c}\oplus\mathfrak{su}_{2,L}\oplus\mathfrak{u}_{1,X}\oplus \mathfrak{u}_{1,Z} \rightarrow \mathfrak{su}_{3,c}\oplus \mathfrak{su}_{2,L}\oplus \mathfrak{u}_{1,Y},  \\
\mathbf{16} &=& \overline{\mathbf{5}}_3\oplus\mathbf{10}_1\oplus\mathbf{1}_{-5} \nonumber \\
&=&  (\overline{\mathbf{3}},\mathbf{1})_{3,2}\oplus(\mathbf{1},\mathbf{2})_{3,-3}\oplus(\mathbf{3},\mathbf{2})_{-1,1}\oplus(\overline{\mathbf{3}},\mathbf{1})_{-1,-4}\oplus(\mathbf{1},\mathbf{1})_{-1,6}\oplus(\mathbf{1},\mathbf{1})_{-5,0}, \nonumber \\
 &=&  (\overline{\mathbf{3}},\mathbf{1})_{\frac{1}{3}}\oplus(\mathbf{1},\mathbf{2})_{-\frac{1}{2}}\oplus(\mathbf{3},\mathbf{2})_{\frac{1}{6}}\oplus(\overline{\mathbf{3}},\mathbf{1})_{-\frac{2}{3}}\oplus(\mathbf{1},\mathbf{1})_1\oplus(\mathbf{1},\mathbf{1})_0, \nonumber \\
\mathbf{45} &=& \mathbf{24}_0 \oplus \mathbf{10}_{4}\oplus\overline{\mathbf{10}}_{-4}\oplus\mathbf{1}_0 = (\mathbf{8},\mathbf{1})_{0,0}\oplus(\mathbf{1},\mathbf{3})_{0,0}\oplus(\mathbf{1},\mathbf{1})_{0,0}\oplus(\mathbf{3},\mathbf{2})_{0,-5}\oplus(\overline{\mathbf{3}},\mathbf{2})_{0,5} \nonumber \\
&& \oplus ({\bf1},{\bf1})_{0,0} \oplus ({\bf1},{\bf1})_{4,6} \oplus ({\bf3},{\bf2})_{4,1}\oplus (\overline{{\bf3}},{\bf1})_{-1,-4} \oplus ({\bf1},{\bf1})_{-4,-6} \oplus (\overline{{\bf3}},{\bf2})_{-4,-1}\oplus ({\bf3},{\bf1})_{-4,4} \nonumber \\
&=& (\mathbf{8},\mathbf{1})_{0}\oplus(\mathbf{1},\mathbf{3})_{0}\oplus(\mathbf{1},\mathbf{1})_{0}\oplus(\mathbf{3},\mathbf{2})_{-\frac{5}{6}}\oplus(\overline{\mathbf{3}},\mathbf{2})_{\frac{5}{6}}\oplus ({\bf1},{\bf1})_{0} \oplus ({\bf1},{\bf1})_{1} \nonumber \\
&& \oplus ({\bf3},{\bf2})_{\frac{1}{6}}\oplus (\overline{{\bf3}},{\bf1})_{-\frac{2}{3}} \oplus ({\bf1},{\bf1})_{-1} \oplus (\overline{{\bf3}},{\bf2})_{-\frac{1}{6}}\oplus ({\bf3},{\bf1})_{\frac{2}{3}}, \nonumber \\
{\bf 10} &=& {\bf 5}_2 \oplus \overline{{\bf 5}}_{-2} = ({\bf 3},{\bf1})_{2,-2} \oplus ({\bf1},{\bf2})_{2,3} \oplus (\overline{{\bf 3}},{\bf 1})_{-2,2} \oplus ({\bf 1},{\bf 2})_{-2,-3} \nonumber \\
&=& ({\bf 3},{\bf1})_{-\frac{1}{3}} \oplus ({\bf1},{\bf2})_{\frac{1}{2}} \oplus (\overline{{\bf 3}},{\bf 1})_{\frac{1}{3}} \oplus ({\bf 1},{\bf 2})_{-\frac{1}{2}}.\nonumber
\label{SU5break}
\end{eqnarray}
For standard $SU(5)$ GUT, the $U(1)_Y$ charge $Q_Y$ is proportional to $Q_Z$ (as shown above), while flipped $SU(5)$ GUT finds $Q_Y$ proportional to $Q_X - Q_Z$.
The $\mathfrak{so}_{10}$ GUT allows for either a ${\bf10}$, ${\bf 120}$, or a $\overline{{\bf126}}$ Higgs~\cite{2019Croon}. Here, the ${\bf 10}$ Higgs of $\mathfrak{so}_{10}$ is shown to break to a ${\bf 5}$ Higgs of $\mathfrak{su}_5$.
Singularities from $E_8$ in F-theory have been argued to lead to flipped $SU(5)$ GUTs~\cite{Chen2011}. Earlier investigations of flipped $SU(5)$ found a hidden $SO(10)\times SO(6)$ gauge group~\cite{Antoniadis:1989zy}, which may naturally fit in $SO(16) \subset E_8$. Throughout, we focus on the real form $E_{8(-24)}$ to include noncompact spacetime symmetries with GUT groups. 


Breaking from $\mathfrak{so}_{10}$ to the $\mathfrak{su}_4\oplus\mathfrak{su}_2\oplus\mathfrak{su}_2$ of Pati-Salam GUT gives
\begin{eqnarray}
\mathfrak{so}_{10} &\rightarrow& \mathfrak{su}_4\oplus\mathfrak{su}_{2}\oplus\mathfrak{su}_2 \rightarrow \mathfrak{su}_{4}\oplus\mathfrak{su}_{2}\oplus \mathfrak{u}_{1,R} \rightarrow \mathfrak{su}_{3,c} \oplus \mathfrak{su}_{2,L}\oplus \mathfrak{u}_{1,R} \oplus \mathfrak{u}_{1,B-L}  \nonumber \\
&\rightarrow& \mathfrak{su}_{3,c} \oplus \mathfrak{su}_{2,L} \oplus \mathfrak{u}_{1,Y}, \nonumber \\
\mathbf{16} &=& (\mathbf{4},\mathbf{2},\mathbf{1}) \oplus (\overline{\mathbf{4}},\mathbf{1},\mathbf{2}) = ({\bf 4},{\bf 2})_{0} \oplus (\overline{{\bf 4}},{\bf 1})_{1} \oplus (\overline{{\bf 4}},{\bf 1})_{-1}  \\
&=& ({\bf3},{\bf2})_{0,-1} \oplus ({\bf1},{\bf2})_{0,3} \oplus (\overline{{\bf 3}},{\bf 1})_{-1,1} \oplus ({\bf 1},{\bf1})_{-1,-3} \oplus (\overline{{\bf3}},{\bf1})_{1,1} \oplus ({\bf1},{\bf1})_{1,-3} \nonumber \\
&=& ({\bf3},{\bf2})_{\frac{1}{6}} \oplus ({\bf1},{\bf2})_{-\frac{1}{2}} \oplus (\overline{{\bf3}},{\bf1})_{-\frac{2}{3}} \oplus ({\bf1},{\bf1})_{0} \oplus (\overline{{\bf3}},{\bf1})_{\frac{1}{3}} \oplus ({\bf1},{\bf1})_{1}, \nonumber \\
\mathbf{45} &=& (\mathbf{15},\mathbf{1},\mathbf{1}) \oplus (\mathbf{1},\mathbf{3},\mathbf{1}) \oplus (\mathbf{1},\mathbf{1},\mathbf{3}) \oplus ({\bf 6},{\bf 2},{\bf 2}) \nonumber \\
&=& ({\bf15},{\bf1})_0 \oplus ({\bf1},{\bf3})_0 \oplus ({\bf1},{\bf1})_0 \oplus ({\bf1},{\bf1})_2 \oplus ({\bf1},{\bf1})_{-2} \oplus ({\bf6},{\bf2})_1 \oplus ({\bf6},{\bf2})_{-1} \nonumber \\
&=& ({\bf8},{\bf1})_{0,0} \oplus ({\bf1},{\bf1})_{0,0} \oplus ({\bf3},{\bf1})_{0,-4} \oplus (\overline{{\bf3}},{\bf1})_{0,4} \oplus ({\bf1},{\bf3})_{0,0} \oplus ({\bf1},{\bf1})_{0,0} \nonumber \\
&& \oplus ({\bf1},{\bf1})_{2,0} \oplus({\bf1},{\bf1})_{-2,0} \ ({\bf3},{\bf2})_{1,2} \oplus (\overline{{\bf3}},{\bf2})_{1,-2} \oplus ({\bf3},{\bf2})_{-1,2} \oplus (\overline{{\bf3}},{\bf2})_{-1,-2}\nonumber \\
&=& ({\bf8},{\bf1})_0 \oplus ({\bf1},{\bf3})_0 \oplus ({\bf1},{\bf1})_0 \oplus ({\bf3},{\bf1})_{\frac{2}{3}} \oplus (\overline{{\bf3}},{\bf1})_{-\frac{2}{3}} \oplus ({\bf 1},{\bf1})_0 \nonumber \\
&& \oplus ({\bf1},{\bf1})_1 \oplus ({\bf 1},{\bf1})_{-1} \oplus ({\bf3},{\bf2})_{\frac{1}{6}} \oplus (\overline{{\bf3}},{\bf2})_{\frac{5}{6}} \oplus ({\bf 3},{\bf2})_{-\frac{5}{6}} \oplus (\overline{{\bf3}},{\bf2})_{-\frac{1}{6}}, \nonumber \\
{\bf 10} &=& ({\bf6},{\bf1},{\bf1}) \oplus ({\bf1},{\bf2},{\bf2}) = ({\bf6},{\bf1})_0 \oplus ({\bf 1},{\bf 2})_{1} \oplus ({\bf 1},{\bf 2})_{-1} \nonumber \\
&=& ({\bf 3},{\bf1})_{0,2} \oplus (\overline{{\bf3}},{\bf 1})_{0,-2} \oplus ({\bf 1},{\bf 2})_{1,0} \oplus ({\bf 1},{\bf 2})_{-1,0} \nonumber \\
&=& ({\bf 3},{\bf1})_{-\frac{1}{3}} \oplus (\overline{{\bf3}},{\bf 1})_{\frac{1}{3}} \oplus ({\bf 1},{\bf 2})_{\frac{1}{2}} \oplus ({\bf 1},{\bf 2})_{-\frac{1}{2}}, \nonumber
\end{eqnarray}
where two algebras $\mathfrak{u}_{1,R}$ and $\mathfrak{u}_{1,B-L}$ can be mixed to give $Q_Y$ proportional to $Q_R + \frac{1}{2}Q_{B-L}$. The electroweak Higgs comes from $({\bf1},{\bf2},{\bf2})$, which fits into the ${\bf 10}$ of $\mathfrak{so}_{10}$. Note that other VEVs are also required for each symmetry breaking, such as $({\bf 15},{\bf1},{\bf1})$ for $\mathfrak{su}_4 \rightarrow \mathfrak{su}_{3,c}\oplus \mathfrak{u}_{1,B-L}$, $({\bf1},{\bf1},{\bf3})$ for $\mathfrak{su}_{2,R}\rightarrow \mathfrak{u}_{1,R}$, and $({\bf4},{\bf1},{\bf2})$ for $\mathfrak{u}_{1,R}\oplus\mathfrak{u}_{1,B-L} \rightarrow \mathfrak{u}_{1,Y}$~\cite{Hartmann:2014fya}. In our interpretation, we will find the that the off-shell fermionic dofs allow for the symmetry breaking to $\mathfrak{u}_{1,Y}$.

The flipped $Spin(10)$ GUT uses the algebra $\mathfrak{so}_{10}\oplus\mathfrak{u}_1$, which is a maximal subalgebra of $\mathfrak{e}_6$ \cite{Maekawa:2003wm,Bertolini:2010yz}. Moreover, $E_6$ GUT uses the entire ${\bf27}$ for describing fermions \cite{Gursey1980}, which is mathematically similar yet physically distinct from other recent work \cite{Dubois-Violette:2016kzx,Todorov:2018mwd,Dubois-Violette:2018wgs,Schwichtenberg:2017xhv}. We rule out this possibility with $\mathfrak{e}_{8(-24)}$, as only ${\bf16}$ of the ${\bf27}$ dofs are fermionic, which will become more clear below.
$Spin(10)$ GUT does not provide any mechanism for naturally describing three generations.

There are three conjugacy classes of $\mathfrak{so}_{10}\oplus\mathfrak{u}_1$ subalgebras in $\mathfrak{e}_6$, related by inner automorphisms of $\mathfrak{e}_6$ itself (in turn related to $\mathfrak{so}_{8}$ triality). Breaking $\mathfrak{e}_6$ to $\mathfrak{so}_{10}\oplus\mathfrak{u}_1$ gives the following branchings,
\begin{eqnarray}
{\bf78} &\rightarrow& {\bf45}_0 \oplus {\bf1}_0 \oplus {\bf16}_{-3} \oplus \overline{{\bf16}}_3, \nonumber \\
{\bf27} &\rightarrow& {\bf16}_1 \oplus {\bf10}_{-2}\oplus {\bf1}_4,
\label{e6break}
\end{eqnarray}
where ${\bf 27}$ and ${\bf78}$ are the fundamental and adjoint of $\mathfrak{e}_6$, respectively.
%
%
The trinification GUT uses $\mathfrak{su}_3\oplus\mathfrak{su}_3\oplus\mathfrak{su}_3$, a maximal and non-symmetric subalgebra of $\mathfrak{e}_6$~\cite{trinification,Babu:1985gi,Lazarides:1993uw,Lazarides:1994px,Kim:2003cha,Willenbrock:2003ca,Carone:2004rp,Hetzel:2015cca,Camargo-Molina:2016bwm,Babu:2017xlu}. Also, $\mathfrak{e}_6$ contains $\mathfrak{su}_6\oplus \mathfrak{su}_2$, which can give $\mathfrak{su}_6$ GUT~\cite{Fukugita:1981gn,Majumdar:1982sx,Tabata:1983cr,PhysRevD.71.095013} with an additional $\mathfrak{su}_{2}$~\cite{Huang:2017uli}. Next, $\mathfrak{e}_7$ and $\mathfrak{e}_8$ are shown to encapsulate $\mathfrak{e}_6$ GUT.


\section{Three mass generations from 6D spacetime}
\label{3massGen}

\subsection{Intuition from the magic star of $\mathfrak{e}_8$}

\begin{figure}
\begin{centering}
\includegraphics[scale=.5]{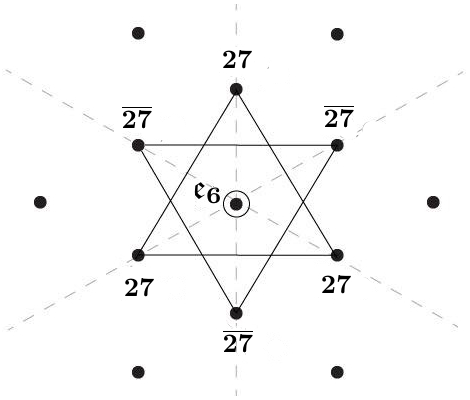}
\caption{The exceptional Lie algebra $\mathfrak{e}_8$ is projected onto a $\mathfrak{g}_2$-like root lattice, which places $\mathfrak{e}_6$ in the center with three fundamental and three anti-fundamental representations of $\mathfrak{e}_6$, ${\bf27}$ and $\overline{{\bf27}}$, respectively.}
\label{e8magic}
\end{centering}
\end{figure}

The only exceptional GUT algebra is $\mathfrak{e}_6$. However, $\mathfrak{e}_7$ and $\mathfrak{e}_8$ contain representations of $\mathfrak{e}_6$ GUT.
The adjoint representation of $E_7$ contains bosons and a single generation of fermions via
\begin{eqnarray}
\mathfrak{e}_7 &\rightarrow& \mathfrak{e}_6\oplus\mathfrak{u}_1, \\
\mathbf{133} &=& \mathbf{78} \oplus \mathbf{1} \oplus \mathbf{27} \oplus \overline{\mathbf{27}}. \nonumber
\end{eqnarray}
From the perspective of GUTs, the utility of $E_7$ is not to generalize $E_6$ GUT to $E_7$ GUT, but to simply place all of the content of $E_6$ GUT for one generation within $E_7$.
The so-called magic star projection of $\mathfrak{e}_8$~\cite{Truini:2011np} breaks to the maximal subalgebra $\mathfrak{su}_3\oplus \mathfrak{e}_{6}$ to naturally give three generations,
\begin{eqnarray}
\mathfrak{e}_8 &\rightarrow& \mathfrak{e}_6 \oplus \mathfrak{su}_3, \\ 
\mathbf{248} &=& (\mathbf{78},\mathbf{1}) \oplus (\mathbf{1},\mathbf{8}) \oplus (\mathbf{27},\mathbf{3}) \oplus (\overline{\mathbf{27}},\overline{\mathbf{3}}), \nonumber
\end{eqnarray}
which can be visualized in Fig.~\eqref{e8magic}. As hinted by the magic star of $\mathfrak{e}_8$ itself, three distinct embeddings of $\mathfrak{e}_7$ are within $\mathfrak{e}_8$ and overlap by $\mathfrak{e}_6$ to give three generations. Breaking $\mathfrak{e}_8\rightarrow \mathfrak{e}_7$ gives the five-grading of contact type,
\begin{eqnarray}
\mathfrak{e}_8 &\rightarrow& \mathfrak{e}_7 \oplus \mathfrak{su}_2 \rightarrow \mathfrak{e}_7 \oplus \mathfrak{u}_1, \\
\mathbf{248} &=& (\mathbf{133},\mathbf{1}) \oplus (\mathbf{1},\mathbf{3}) \oplus (\mathbf{56},\mathbf{2}) = \mathbf{1}_{-2} \oplus \mathbf{56}_{-1} \oplus (\mathbf{133}_0 \oplus \mathbf{1}_0) \oplus \mathbf{56}_1 \oplus \mathbf{1}_2. \nonumber
\end{eqnarray}

Unsurprisingly, $\mathfrak{e}_6\oplus \mathfrak{su}_3$ has been embedded inside $\mathfrak{e}_8$ to give a way to extend $\mathfrak{e}_6$ GUT to include a family unification $\mathfrak{su}_{3,F}$~\cite{Barr:1987pu,Kawase:2010na}. Also, $\mathfrak{e}_8$ contains $\mathfrak{su}_3\oplus\mathfrak{su}_3\oplus\mathfrak{su}_3\oplus\mathfrak{su}_3$, suggesting a quadrification model as trinification with a global family $\mathfrak{su}_{3,F}$~\cite{Camargo-Molina:2016bwm}. However, we take a different approach in this work. The only real form of $\mathfrak{e}_8$ that has a chance of obtaining $\mathfrak{so}_{10}$ and $\mathfrak{so}_{3,1}$ is $\mathfrak{e}_{8(-24)}$. While this can't give $\mathfrak{su}_{3,F}$, $\mathfrak{so}_{3,F}$ is suggested via three timelike dimensions to motivate $\mathfrak{so}_{3,3}$ spinors. The $\mathfrak{e}_{8(-24)}$ algebra is natural for three generations of matter that are efficiently encoded in 128 off-shell dofs.

The magic star projection of $\mathfrak{e}_8$ in Fig.~\eqref{e8magic}~\cite{Truini:2017jiy} isolates six exceptional Jordan algebras surrounding $\mathfrak{e}_6$\footnote{Strictly speaking, only the noncompact real forms $\mathfrak{e}_{6(-26)}$ and $\mathfrak{e}_{6(6)}$ admit real forms of exceptional Jordan algebras as fundamental representations.
}. The Peirce decomposition of the exceptional Jordan algebra (cfr. e.g. Chap. 8 of \cite{PD}) occurs when breaking from $\mathfrak{e}_6$ to $\mathfrak{so}_{10}$; as shown in Eq.~\eqref{e6break},
three $\mathbf{16}$'s and three $\overline{\mathbf{16}}$'s emerge from $\mathfrak{e}_8$, which are representations of $\mathfrak{so}_{10}$ and give the 96 on-shell fermionic dofs. Following the breaking of $\mathfrak{e}_6 \rightarrow \mathfrak{so}_{10}$ shown in Eq.~\eqref{e6break}, an additional $\mathbf{16}\oplus \overline{\mathbf{16}}$ is contained within the adjoint of $\mathfrak{e}_6$ (in the center of the magic star in Fig.~\eqref{e8magic}). The $\mathbf{16}\oplus \overline{\mathbf{16}}$ inside $\mathfrak{e}_6$ represent additional off-shell fermionic dofs, as $\mathfrak{e}_8$ contains $128$ spinorial roots, 96 of which are outside of $\mathfrak{e}_6$. As mentioned above, the magic star of $\mathfrak{e}_8$, shown in Fig.~\eqref{e8magic}, allows for three embeddings of $\mathfrak{e}_7$ within $\mathfrak{e}_8$, and each of them contain the same central $\mathfrak{e}_6$. Thus, the magic star projection of $\mathfrak{e}_8$ provides a simple geometric way to see how three mass generations of fermions fit inside the ${\bf128}$ spinor representation inside $\mathfrak{e}_8$.

It has been stated that all GUTs besides $\mathfrak{su}_5$ and $\mathfrak{so}_{10}$ require mirror fermions~\cite{Maalampi:1988va}, unless supersymmetry is introduced. Mirror fermions must have weak hypercharge for the right-handed chirality states, instead of the left-handed. The relationship between three generations and mirror fermions is discussed in the next section.

\subsection{Three momenta with different mass}

In this section, $\mathfrak{so}_{3,3}$ spacetime is shown to efficiently encode three Dirac spinors of different mass in 16 off-shell dofs, provided that they have the same charge. A 4D Dirac spinor is given by 8 off-shell dofs, implying that three generations of a particle requires 24 off-shell dofs. The scattering amplitudes community utilizes massless $\mathfrak{so}_{5,1}$ spinors to encode massive 4D spinors for computational simplicity~\cite{Cheung:2009dc}. Other work also utilized $\mathfrak{so}_{5,1}$ or six dimensions for three generations~\cite{Hall:2001xr,Babu:2002ti}. Spacetime from $\mathfrak{so}_{3,3}$ more clearly allows for three $\mathfrak{so}_{3,1}$ spacetimes as a subalgebra. 
The intuition for multiple time dimensions is to encode multiple mass generations. While $\mathfrak{so}_1$ for a single time is a trivial algebra with no generators, $\mathfrak{so}_3$ allows for an explanation for three generations~\cite{Wilczek:1981iz,BenTov:2015gra,Reig:2017nrz,Reig:2018ocz}, which we explore here as a spacetime symmetry that generalizes the Lorentz and conformal algebras with extra time dimensions.

To briefly show how $\mathfrak{so}_{3,3}$ encodes three masses, consider a 6D massive vector $p^{\bar{\mu}}$, where $\bar{\mu} = -3,-2,-1,1,2,3$. It is clear that we can find three 4D momenta $p_i^\mu$ for $i=1,2,$ or $3$, where time is taken from $\bar{\mu}=-1, -2$, or $-3$. Given that $p_{\bar{\mu}}p^{\bar{\mu}} = -m_6^2$ with a positive signature for space,
\begin{eqnarray}
m_1^2 &=& - p_{1,\mu}p_1^\mu = m_6^2 - p_{-3}^2 - p_{-2}^2, \nonumber \\
m_2^2 &=& - p_{2,\mu}p_2^\mu = m_6^2 - p_{-3}^2 - p_{-1}^2, \\
m_3^2 &=& - p_{3,\mu}p_3^\mu = m_6^2 - p_{-2}^2 - p_{-1}^2. \nonumber
\end{eqnarray}
This demonstrates that a 6D momentum can be projected to three generations of 4D momenta with different masses $m_1, m_2$, and $m_3$. To obtain 4D spinors, each momentum $p_i^\mu$ can be decomposed into spinors via the isomorphism $\mathfrak{so}_{3,1}\sim \mathfrak{sl}_{2,\mathbb{C}}$.
We consider $\mathfrak{so}_{4,4}$ as the conformal group of $\mathfrak{so}_{3,3}$. The three extra time dimensions in $\mathfrak{so}_{4,4}$ provide a geometrical origin of comprehensive family unification proposed by Wilczek et al.~\cite{Wilczek:1981iz,BenTov:2015gra,Reig:2017nrz,Reig:2018ocz}. Ghosts are often thought to make multiple time dimensions problematic, but here, focus is given to spinors in $D=3+1$ taken from representations of $\mathfrak{so}_{3,3}$ and twistor representations of $\mathfrak{so}_{4,4}$.


Before diving into $\mathfrak{e}_{8(-24)}$ and the eight charges associated with fermions found in the SM, we demonstrate how $\mathfrak{f}_{4(4)}$ can be used to efficiently contain spacetime symmetry with fermions and anti-fermions with their mirrors in a single algebra. The fermions are contained in the ${\bf 128}$ of $\mathfrak{so}_{4,12}$ inside $\mathfrak{e}_{8(-24)}$. The maximal and non-symmetric subalgebra $\mathfrak{f}_{4(4)}\oplus \mathfrak{g}_{2(-14)}$ is found within $\mathfrak{e}_{8(-24)}$ \cite{dGM} to give the following representations,
\begin{eqnarray}
\mathfrak{e}_{8(-24)} &\rightarrow& \mathfrak{f}_{4(4)}\oplus \mathfrak{g}_{2(-14)}, \\
{\bf248} &=& ({\bf52},{\bf1}) \oplus ({\bf1},{\bf14}) \oplus ({\bf 26},{\bf7}). \nonumber
\end{eqnarray}
Note that $\mathfrak{f}_{4(4)}$ contains $\mathfrak{so}_{4,5}$ as a maximal (and symmetric) subalgebra. The ${\bf16}$ of $\mathfrak{so}_{4,5}$ can be found inside the ${\bf26}$ of $\mathfrak{f}_{4(4)}$. Since $\mathfrak{f}_{4(4)}$ also contains a ${\bf 16}$, the fermions as ${\bf128}$ of $\mathfrak{so}_{4,12}$ are contained in one ${\bf16}$ inside $\mathfrak{f}_{4(4)}$ and seven ${\bf16}$'s inside $({\bf 26},{\bf7})$ of $\mathfrak{f}_{4(4)}\oplus\mathfrak{g}_{2(-14)}$. The SM contains eight charge configurations of the electron, three up quarks, three down quarks, and neutrino. Focusing on $\mathfrak{f}_{4(4)}$ allows for the isolation of a singe charge configuration, such as the electron, giving ${\bf16}$ instead of ${\bf 128}$. As shown above, the $\mathfrak{so}_{3,3}$ inside $\mathfrak{f}_{4(4)}$ allows for three distinct $\mathfrak{so}_{3,1}$ spacetime algebras with different timelike projections that can lead to different masses in 4D spacetime.

As hinted at above, focusing on $\mathfrak{f}_{4(4)}$ allows for a simple demonstration of how three generations of fermions are contained inside $\mathfrak{e}_{8(-24)}$, before diving into all of the different charge configurations. The $\mathfrak{f}_{4(4)}$ algebra is broken in the following manner:
\begin{eqnarray}
\mathfrak{f}_{4(4)} &\rightarrow& \mathfrak{so}_{4,5} \rightarrow \mathfrak{so}_{4,4} \rightarrow \mathfrak{so}_{3,3}\oplus\mathfrak{so}_{1,1}\rightarrow \mathfrak{so}_{3,1}\oplus\mathfrak{so}_{1,1}\oplus\mathfrak{u}_{1}. \nonumber \\
{\bf 52} &=& {\bf 36} \oplus {\bf 16} = {\bf 28} \oplus {\bf 8}_v \oplus {\bf 8}_s \oplus {\bf 8}_c \\
&=& {\bf 15}_0 \oplus {\bf 1}_0 \oplus {\bf 6}_2 \oplus {\bf 6}_{-2} \oplus {\bf 6}_0 \oplus {\bf 1}_2 \oplus {\bf 1}_{-2} \oplus {\bf 4}_1 \oplus {\bf 4}'_{-1} \oplus {\bf 4}'_1 \oplus {\bf 4}_{-1} \nonumber \\
&=& ({\bf 3},{\bf 1})_{0,0} \oplus ({\bf 1},{\bf 3})_{0,0} \oplus ({\bf 1},{\bf 1})_{0,0} \oplus ({\bf 2},{\bf 2})_{0,2} \oplus ({\bf 2},{\bf 2})_{0,-2} \oplus ({\bf 1},{\bf 1})_{0,0} \nonumber \\
&& \oplus ({\bf 2},{\bf 2})_{0,0} \oplus ({\bf 1},{\bf 1})_{0,2} \oplus ({\bf 1},{\bf 1})_{0,-2} \nonumber \oplus ({\bf 2},{\bf 2})_{2,0} \oplus ({\bf 1},{\bf 1})_{2,2} \oplus ({\bf 1},{\bf 1})_{2,-2} \nonumber \\
&& \oplus ({\bf 2},{\bf 2})_{-2,0} \oplus ({\bf 1},{\bf 1})_{-2,2} \oplus ({\bf 1},{\bf 1})_{-2,-2} \oplus ({\bf 1},{\bf1})_{2,0} \oplus ({\bf 1},{\bf 1})_{-2,0} \nonumber \\
&& \oplus ({\bf 2},{\bf 1})_{1,1} \oplus ({\bf 1},{\bf 2})_{1,-1} \oplus ({\bf 1},{\bf 2})_{-1,1} \oplus ({\bf 2},{\bf 1})_{-1,-1} \nonumber \\
&& \oplus ({\bf 1},{\bf 2})_{1,1} \oplus ({\bf 2},{\bf 1})_{1,-1} \oplus ({\bf 2},{\bf 1})_{-1,1} \oplus ({\bf 1},{\bf 2})_{-1,-1}. \nonumber
\label{f44}
\end{eqnarray}
Mirror fermions are identified when breaking $\mathfrak{so}_{3,3}$ spacetime to $\mathfrak{so}_{3,1}\oplus\mathfrak{u}_1$ and thinking of the $\mathfrak{u}_1$ as a dummy (electric) charge (cf. the second subscript in the last step of ~\eqref{f44}). The weight of $\mathfrak{so}_{1,1}$ (cf. the first subscript in the last step of ~\eqref{f44}) identifies the mirror fermions with a $-1$. Spinors of $\mathfrak{so}_{3,3}$ combine the fermion of one chirality with the anti-fermion of the opposite chirality.

While not experimentally measured, mirror fermions preserve symmetry with the weak force, which must acquire a large mass if physical. Mirror fermions typically require additional fermionic dofs. Given that we will use $\mathfrak{e}_{8(-24)}$ to give three generations of matter in 128 dofs, the mirror fermions in $\mathfrak{so}_{3,1}$ spacetime for a single generation are created from the on-shell dofs from the other generations in $\mathfrak{so}_{3,3}$, not off-shell dofs. This subtlety leads to 3 generations in 96 on-shell dofs without propagating mirror fermions from 128 off-shell dofs, not 192 as needed with $D=3+1$ spinors. It is clear that we should not assign $\mathfrak{e}_8$ roots to dofs at the beginning, but rather symmetry break and see what particles arise at lower energy phases. In generalizing to $\mathfrak{e}_{8(-24)}$, the $\mathfrak{u}_1$ in Eq.~\eqref{f44} will be replaced by $\mathfrak{so}_{10}$. The $\mathfrak{u}_1$ can be thought as providing a charge, which helps identify fermions vs antifermions.

To explicitly demonstrate that three generations of a single massless Dirac spinor can be encoded in 16 off-shell degrees of freedom instead of 24, consider the ${\bf16}$ representation inside $\mathfrak{f}_{4(4)}$ as a Majorana spinor $\Psi\in \mathbb{R}^{16}$ of $D=4+4$. Two Majorana-Weyl spinors can be combined to make a single Majorana spinor in $D=3+3$ and $D=4+4$. In this manner, the ${\bf16}={\bf8}_s \oplus {\bf8}_c$ spinor in $D=4+4$ is an $\mathcal{N}=(1,1)$ spinor representation with two chiralities (if studied as a supermultiplet). The Clifford algebra $Cl(4,4)$ leads to a set of $16\times 16$ matrices. Next, we work towards an explicit set of projection matrices for three generations.

Our chosen basis for $Cl(4,4)$ is generated by recursively taking tensor products of $Cl(1,1)$ \cite{Floerchinger}. Each set of four $Cl(1,1)$'s are spanned by two elements $e_{-i}$ and $e_i$ with $i=1,2,3,4$. The signature is encoded in the index, since $e_{-i}^2 = -1$ and $e_i^2 = 1$. The matrix representation of $Cl(1,1)$ is given by $2\times 2$ matrices,
\begin{equation}
e_{-i} = \left(\begin{array}{cc} 0 & -1 \\ 1 & 0 \end{array}\right), \qquad e_i = \left(\begin{array}{cc} 0 & 1 \\ 1 & 0 \end{array} \right).
\end{equation}
Our basis of $16\times 16$ matrices can then be written as 
\begin{eqnarray}
\Gamma^{-3} &=& 1_{2\times 2}\otimes 1_{2\times 2}\otimes 1_{2\times 2}\otimes e_{-4} \nonumber \\ 
\Gamma^{-2} &=& 1_{2\times 2}\otimes 1_{2\times 2}\otimes e_{-3} \otimes e_{-44} \nonumber \\ 
\Gamma^{-1} &=& 1_{2\times 2}\otimes e_{-2} \otimes e_{-33}\otimes e_{-44} \nonumber \\ 
\Gamma^{0} &=& e_{-1}\otimes e_{-22}\otimes e_{-33} \otimes e_{-44} \nonumber \\ 
\Gamma^{1} &=& e_1 \otimes e_{-22}\otimes e_{-33}\otimes e_{-44} \\ 
\Gamma^{2} &=& 1_{2\times 2}\otimes e_2 \otimes e_{-33}\otimes e_{-44} \nonumber \\ 
\Gamma^{3} &=& 1_{2\times 2}\otimes 1_{2\times 2}\otimes e_3 \otimes e_{-44} \nonumber \\ 
\Gamma^{4} &=& 1_{2\times 2}\otimes 1_{2\times 2}\otimes 1_{2\times 2}\otimes e_4 , \nonumber
\end{eqnarray}
where $e_{-ii} = e_{-i}e_i$ for $i=1,2,3,4$.

The bivectors of $Cl(4,4)$ act as generators of $Spin(4,4)$, which can be thought of as a conformal symmetry for $D=3+3$. For de Sitter spacetime with an $S^2$ for mass-flavor oscillations, the following isometry groups are found,
\begin{equation}
Isom(dS_4)\times Isom(S^2) = Spin(4,1) \times Spin(3),
\end{equation}
where $dS_4 = Spin(4,1)/Spin(3,1)$ and $S^2 = Spin(3)/Spin(2)$ provide quotient space realizations. 
Interpreting $Spin(3)$ as a flavor symmetry leads to a way to naturally isolate three generations from a single spinor of $Cl(4,4)$. To find chiral projection operators with respect to $D=3+1$, the emergence of imaginary units must be accounted for. Fortunately, the three extra time dimensions admit three bivectors that are isomorphic to quaternionic imaginary units that implement $Spin(3)_F$ mass/flavor rotations,
\begin{eqnarray}
i &\equiv& \Gamma^{-2-3}= 1_{2\times 2} \otimes 1_{2\times 2} \otimes e_{-3} \otimes e_4 ,\nonumber \\ 
j &\equiv& = \Gamma^{-3-1} = 1_{2\times 2} \otimes e_{-2} \otimes e_{-33} \otimes -e_4,\\ 
k &\equiv& = \Gamma^{-1-2} = 1_{2\times 2} \otimes e_{-2} \otimes e_3 \otimes 1_{2\times 2}, \nonumber
\end{eqnarray}
where $ijk = -1$. 
By focusing on the $Cl(3,4)$ subsector of $Cl(4,4)$ and removing the extra spatial dimension, three different $D=3+1$ chiral projection operators $P_{\pm,i}^{NM}$ can be identified by generalizing the notion of $\gamma^5$,
\begin{eqnarray}
P_{\pm,1}^{NM} &=& \frac{1}{2}\left( 1_{16\times 16} \pm i \Gamma^{0123}\right) = \frac{1}{2}\left( 1_{16\times 16} \pm \Gamma^{-2-30123}\right), \nonumber \\ 
P_{\pm,2}^{NM} &=& \frac{1}{2}\left( 1_{16\times 16} \pm j \Gamma^{0123}\right) = \frac{1}{2}\left( 1_{16\times 16} \pm \Gamma^{-3-10123}\right), \\ 
P_{\pm,3}^{NM} &=& \frac{1}{2}\left( 1_{16\times 16} \pm k \Gamma^{0123}\right) = \frac{1}{2}\left( 1_{16\times 16} \pm \Gamma^{-1-20123}\right). \nonumber
\end{eqnarray}
where $P_{+,u}$ and $P_{-,i}$ for $i=1,2,3$ correspond to the right- and left-chiral projection operators, respectively, and $\Gamma^{-2-30123} = \Gamma^{-2}\Gamma^{-3}\Gamma^0\Gamma^1\Gamma^2\Gamma^3$, similarly for other cases. These do not project out the mirror fermionic states; it should not be possible to find $4\times 3\times 2$ on-shell dofs in 16 off-shell dofs. To project to the normal matter ($N$) and remove the mirror matter ($M$), the representation theory implies that projectors for $Spin(1,1)$ weights is needed. This leads to three sets of $D=3+1$ Dirac spinors from the projectors $P_{i}^{N}$,
\begin{eqnarray}
P_1^N &=& \frac{1}{2}\left(1_{16\times 16} + \Gamma^{-14}\right), \nonumber \\ 
P_2^N &=& \frac{1}{2}\left(1_{16\times 16} + \Gamma^{-24}\right), \\ 
P_3^N &=& \frac{1}{2}\left(1_{16\times 16} + \Gamma^{-34}\right). \nonumber
\end{eqnarray}
Three sets of Dirac spinors embedded in $\mathbb{R}^8$ subsectors of $\mathbb{R}^{16}$ are identified as $\psi_i$,
\begin{equation}
\psi_i = P_i^N \Psi
\end{equation}

Combining both sets of projection operators allows for three sets of two projection operators for three generations of two chiral spinors $P_{\pm,i}^{N}$,
\begin{equation}
P_{\pm,i}^{N} = P_{\pm,i}^{NM} P_i^N,
\end{equation}
where no summation over $i$ is taken above. 
Six sets of chiral spinors can be found via $P_{\pm,i}^{N}\Psi$,
\begin{equation}
\lambda_i = P_{-,i}^N \Psi, \qquad \tilde{\xi}_i = P_{+,i}^N \Psi. \label{chiralProj}
\end{equation}
As seen above, three planes with $Spin(1,1)$ symmetry are spanned by the fourth spatial and one of three extra time dimensions, leading to three independent conformal subspaces found within $Cl(4,4)$. 

The final step to construct a Lagrangian for three generations of fermions from generalizations of Dirac matrices is to construct the analogue of complex conjugation, as $\overline{\psi} = \psi^\dagger \gamma^0$ is a spinor of $\mathbb{C}^4$, while $P_i^N$ projects to three separate spaces of $\mathbb{R}^8 \subset \mathbb{R}^{16}$. Recall that $\mathbb{C}$ can be embedded in $\mathbb{R}^2$, such that $z=a+ib$ is represented as a real vector $(a,b)^\top$. Multiplication by $i$ and complex conjugation are conveniently implemented by generators of $\mathfrak{sl}_{2,\mathbb{R}}$ or elements of $Cl(1,1)$,
\begin{eqnarray}
z^* &=& \sigma_z z = \left(\begin{array}{cc} 1 & 0 \\ 0 & -1 \end{array}\right)\left(\begin{array}{c} a \\ b \end{array}\right) = \left(\begin{array}{c} a \\ -b \end{array}\right), \\ 
iz &=& \sigma_{-iy} z = \left(\begin{array}{cc} 0 & -1 \\ 1 & 0 \end{array}\right) \left(\begin{array}{c} a \\ b \end{array}\right) = \left(\begin{array}{c} -b \\ a \end{array}\right), 
\end{eqnarray}
where $\mathfrak{sl}_{2,\mathbb{R}}$ is spanned by $\sigma_x$, $\sigma_{-iy}$, and $\sigma_z$. 
By recognizing that $i$ admits a $16\times 16$ representation of the $\mathfrak{sl}_{2,\mathbb{R}}$ generator $\sigma_{-iy}$, its realization as a Kronecker product of four $Cl(1,1)$ algebra elements helps identify the appropriate generalization of $\sigma_z$ to implement the analogue of complex conjugation of the three real 8-dimensional spinors $\psi_i$ inside the 16-dimensional spinor $\Psi$. Identifying 3 sets of $16\times 16$ $SL(2,\mathbb{R})$ generators $\Sigma_{x,j}$, $\Sigma_{-iy,j}$, $\Sigma_{z,j}$ with $j=1,2,3$ allows for $\Sigma_{z,i}$ to apply three independent complex conjugations
\begin{equation}
\overline{\psi}_i = \psi_i^\dagger\gamma^0 = \psi_i^{\top *}\gamma^0 = \psi_i^\top\Sigma_{z,i} \Gamma^0,
\end{equation}
where the matrix representations of $\Sigma_{z,i}$ are 
\begin{eqnarray}
\Sigma_{z,1} &=& \Gamma^{4-2} = 1_{2\times 2} \otimes 1_{2\times 2} \otimes e_{-3} \otimes - e_{-4}, \nonumber \\ 
\Sigma_{z,2} &=& \Gamma^{4-3} = 1_{2\times 2} \otimes 1_{2\times 2} \otimes 1_{2\times 2} \otimes e_{4-44}, \\ 
\Sigma_{z,3} &=& \Gamma^{4-1} = 1_{2\times 2} \otimes e_{-2} \otimes e_{3-3} \otimes e_{-4}. \nonumber 
\end{eqnarray}
The three sets of $SL(2,\mathbb{R})$ are contained within $SL(2,\mathbb{C})$ that commutes with the Lorentz group of spacetime and is contained and realized within $Spin(4,4)$ as bivectors of $Cl(4,4)$. 

An explicit Lagrangian in terms of 16-dimensional real non-chiral spinors of $Cl(4,4)$ for 3 massless Dirac spinors is obtained by 
\begin{eqnarray}
\mathcal{L} &=& \overline{\psi}_1 i\gamma^\mu \partial_\mu \psi_1 + \overline{\psi}_2 j\gamma^\mu \partial_\mu \psi_2 + \overline{\psi}_3 k\gamma^\mu \partial_\mu \psi_3 \nonumber \\
&=& \Psi^\top \left(\sum_{i=1}^3 P_i^{N\top} \Sigma_{z,i}\Gamma^0 \Gamma^{-i}\Gamma^{-3-2-1} \Gamma^\mu \partial_\mu \right) \Psi.
\end{eqnarray}
Similarly, the chiral projection operators in Eq.~\eqref{chiralProj} can lead to three generations of Weyl spinors. 


When generalizing from $F_{4(4)}$ to $E_{8(-24)}$, the non-chiral Majorana ${\bf 16}$ spinor of $Spin(4,5)$ is replaced with a chiral Majorana-Weyl ${\bf128}$ spinor of $Spin(4,12)$, which contains three sets of off-shell ${\bf 64}$ spinors with ${\bf32}$ on-shell degrees of freedom that are independent. To construct a Lagrangian for three generations of eight independent fermions, generators of $Cl(3,11)$ can be identified to give $128 \times 128$ matrices, giving a single Majorana spinor in $D=3+11$, which stems from a Majorana-Weyl spinor in $D=4+12$. To generalize $Cl(4,4)$, generators for $Cl(4,12)$ are identified as $256\times 256$ matrices, where the chirality operator can be identified to project to a $128$-dimensional subspace for a single Majorana-Weyl spinor.

One generation of the standard model with Dirac neutrinos can be described by 16 chiral spinors, giving 64 off-shell degrees of freedom. By uplifting to $D=3+3$ spinors, three generations can be found in 128 off-shell degrees of freedom by applying $P_{\pm,i}^{N}$ to 16 sets of Majorana spinors in $D=3+3$. The action for kinetic terms of three generations can be written as 
\begin{eqnarray}
S_{kin} &=& \int d^4x \sum_i^3\left(q^\dagger_i \Gamma^{-i}\Gamma^{-3-2-1} \bar{\sigma}^\mu \partial_\mu q_i + u^\dagger_i \Gamma^{-i}\Gamma^{-3-2-1} \bar{\sigma}^\mu \partial_\mu u_i + d^\dagger_i \Gamma^{-i}\Gamma^{-3-2-1} \bar{\sigma}^\mu \partial_\mu d_i \right. \nonumber \\
&& \left. + l^\dagger_i \Gamma^{-i}\Gamma^{-3-2-1} \bar{\sigma}^\mu \partial_\mu l_i + e^\dagger_i \Gamma^{-i}\Gamma^{-3-2-1} \bar{\sigma}^\mu \partial_\mu e_i + \nu^\dagger_i \Gamma^{-i}\Gamma^{-3-2-1} \bar{\sigma}^\mu \partial_\mu \nu_i\right),
\end{eqnarray}
where $\bar{\sigma}^\mu$ are the identity matrix combined with -1 times three Pauli matrices.
While the standard model is often formulated from left-chiral spinors only, this is equivalent to our convention of choosing right-chiral spinors for weak isospin singlets $u_i$, $d_i$, $e_i$, and $\nu_i$ found from $P_{-,i}^N$ acting on four sets of ${\bf16}$ spinors $U, D, E, N \in \mathbb{R}^{16}$ as spinors of $Spin(4,4)$. The weak isospin doublets $q_i$ and $l_i$ are taken as left-chiral spinors found from $P_{-,i}^N$ acting on the same spinors $U$, $D$, $E$, and $N$, where $q_i$ contains spinors from $U$ and $D$, while $l_i$ contains spinors from $E$ and $N$. Note that there are three sets of $U$ and $D$ for three colors (indices suppressed), resulting in 8 copies of 16-spinors in total. 
Interaction terms are found by uplifting partial derivatives to covariant derivatives for flavor eigenstates based on charges found from representation theory. 

By identifying the three sets of complex structures embedded in real vectors, the following electroweak Yukawa interactions with the Higgs field $\Phi$ are shown for mass eigenstates,
\begin{equation}
S_{Yukawa} = \int d^4x \left(-Y_u^{ij} q^\dagger_i \Phi^*u_j - Y_d^{ij} q_i^\dagger \Phi d_j - Y_\nu^{ij}l_i^\dagger \Phi^* \nu_j - Y_e^{ij} l^\dagger_i \Phi e_j  + \mbox{h.c.}\right),
\end{equation}
where $u_j$, $d_j$, $\nu_j$, and $e_j$ are mass eigenstates of weak isospin singlets in terms of right-chiral fermions found with $P_{+,i}^N$, while $q_i$ and $l_i$ are left-chiral weak isospin doublets found from $P_{-,i}^N$. The $q_i^\dagger$ spinor takes a transpose, yet must also implement the analogue of complex conjugation with $\Sigma_{z,i}$. $\Sigma_{z,i}$ is also used for finding the analogue of the Hermitian conjugate term. The mass eigenvalues can be found as eigenvalues of the matrices $Y_f^{ij}\left<\Phi\right>$. The three complex subspaces overlap to give $Spin(3)$ as imaginary quaternions for mass-flavor oscillations and make contact with $SO(3)$ models such as Refs.~\cite{Wilczek:1981iz,King:2005bj,BenTov:2015gra,Reig:2017nrz,Reig:2018ocz}. While the Lagrangian uses spinors from $D=3+3$, the global spacetime manifold is restricted to $D=3+1$.

This realization of the quaternions from Clifford algebras for three generations is similar to recent work by Wilson \cite{Wilson}, except here, the quaternionic units emerge naturally from the rotations within $\mathfrak{so}_{3,F}$ as bivectors of $Cl(3,4)$, while Wilson considers $Cl(3,3)$ from $Cl(3,1)_{\mathbb{H}}$. By starting with a 16-dimensional real spinor for $Cl(3,3)$, it is more difficult to identify the irreducible spinor representations of $D=3+3$ and the $Spin(3)$ flavor symmetry to implement mass/flavor oscillations. By projecting on chiral states of normal matter, $Cl(4,4)$ can appropriately project to 6 sets of chiral spinors with 4 off-shell dofs, just as contained in a Weyl spinor of $D=3+1$ and is found in the standard model. 

While it is often thought that complex spinor representations are required for chiral spinors, this is not true whenever Majorana-Weyl spinors are allowed, which occurs for the ${\bf8}_s\oplus{\bf8}_c$ spinors of $\mathfrak{so}_8$ or $\mathfrak{so}_{4,4}$ in $\mathfrak{f}_4$ and ${\bf 128}$ spinors in all real forms of $\mathfrak{e}_8$. The ${\bf16}$ spinor in $\mathfrak{f}_{4(4)}$ can also be represented by a split-octonionic spinor of $\mathbb{O}_s^2$, which clearly admits three complex subsets, since $\mathbb{H}^2 \in \mathbb{O}_s^2$ and the quaternions $\mathbb{H}$ contain three imaginary units. Wilson, Dray, and Manogue have recently explored the octo-octonionic structure of the $\mathfrak{e}_8$ Lie algebra as well as a physics proposal for $\mathfrak{e}_{8(-24)}$ \cite{2022WDM,2022MDW}. While their work focuses on recovering gluon-like fields from the noncompact algebra $\mathfrak{sl}_{3,\mathbb{R}}$, ours focuses on recovering GUTs with standard QCD using the compact $\mathfrak{su}_3$ algebra. The use of split octonions for three generations of matter have been discussed by Gogberashvili \cite{Gogberashvili}.

This section demonstrates that three on-shell generations of Dirac fermions with complex representations can be found from a single off-shell Majorana spinor of $Cl(4,4)$, which can be found as the ${\bf16}$ representation $\mathbb{O}_s\mathbb{P}^2 = F_{4(4)}/Spin(5,4)$. Projection operators for three generations of chiral spinors provide an understanding of the quaternionic subspaces of $\mathbb{O}_s^2$ spinors, leading to three complex subspaces in $\mathbb{H}$ and six in $\mathbb{O}_s$ for two chiralities of each generation. Since the standard model includes eight fermionic charge configurations of the electron, three down quarks, neutrino, and three up quarks, totaling 128 off-shell degrees of freedom contained within $(\mathbb{O}_s \otimes \mathbb{O})\mathbb{P}^2 = E_{8(-24)}/Spin(12,4)$ \cite{2022Marrani}. If the standard model combined with gravity can be found in a single ${\bf248}$ representation of $\mathfrak{e}_8$, it can only be done if Majorana-Weyl spinors of $D=3+3$, $D=4+4$, $D=11+3$, or $D=12+4$ are utilized, thus confirming Distler and Garibaldi's refutation of $Spin(3,1)$ spinors \cite{Distler:2009jt}, yet placing a later conjecture of Lisi's use of $Spin(4,4)\times Spin(8)$ on a more rigorous footing \cite{Lisi:2015oja}. 

Rather than supposing that $Spin(4,4)\times Spin(8)$ gauge theory should be pursued, we propose that this structure or $Spin(3,3)\times Spin(9,1)$ can be viewed as a global symmetry that is dual or hidden from the gauge theoretic structure. In this manner, these results may be helpful for string theory, as supergravity compactifications often lead to $Spin(8)$ gauge symmetry, including the study of $AdS_4\times S^7$ as a compactification of M-theory. By considering $dS_4\times S^2 \times S^7$ as a global spacetime manifold structure, the gauge gravity formulation of Ivanov and Niederle can be employed to find $M \times G$ symmetry with $M$ as a global spacetime manifold and $G$ as a local gauge group \cite{IvanovNiederle}. Our notion of a unified field theory with an internal double copy is one where $M$ and $G$ are treated as different objects at low energies, yet stemming from the same symmetry group at high energies. While this sounds similar to $E_8\times E_8$ heterotic string theory \cite{Green:1984sg,CANDELAS1985}, this proposal is unique in the sense that string theory has only studied $E_{8(-24)}$ in the context of magic supergravity U-dualities \cite{magicSUGRA}, which is different. Instead, this work suggests the exploration of a $D=12+4$ supermembrane theory with a 4-brane and an S-dual 8-brane that contains superalgebras of M-theory, F-theory, and S-theory \cite{Rios:2018lhc}. A membrane realization of the high energy theory containing all of the subsequent representation theory is outside the scope of this work. 


\section{High energy theories from four spacelike dimensions}
\label{4spatialDim}

This section looks to generalize the work of Nesti and Percacci, who used $\mathfrak{so}_{3,11}\oplus {\bf 64}$ to describe $SO(10)$ GUT with spacetime for one generation~\cite{Nesti:2009kk,Percacci:2009ij}. Thus, we will consider the maximal subalgebra $\mathfrak{so}_{4,12}$ of $\mathfrak{e}_{8(-24)}$ to have its 16-dimensional vector representation ${\bf16}$ with signature $(s,t)=(4,12)$, where $s$ and $t$ respectively denote the number of spacelike and timelike dimensions. With the utilization of $\mathfrak{so}_{3,3}$ and the intuitive picture provided by the magic star projection of $\mathfrak{e}_{8(-24)}$ (discussed in the previous section), we look to establish how the SM and spacetime can fit into various high energy theories. Additionally, new routes that directly lead to $SU(5)$ and Pati-Salam GUTs that bypass $\mathfrak{so}_{10}$ are found. The utilization of $\mathfrak{e}_{8(-24)}$ is more similar to a Lie group cosmology model~\cite{Lisi:2015oja} than Ref.~\cite{Lisi:2007gv}, yet we differ on the fermionic interpretations and demonstrate how this noncompact real form connects to various GUTs.

The following breaking could be taken to isolate $\mathfrak{so}_{3,3}$ spacetime,
\begin{equation}
\mathfrak{e}_{8(-24)} \rightarrow \mathfrak{so}_{4,12} \rightarrow \mathfrak{so}_{3,3}\oplus\mathfrak{so}_{1,9}.
\end{equation}
While this breaking isolates a spacetime of interest, it introduces a Lorentzian $\mathfrak{so}_{1,9}$ charge algebra, which is not ideal for connecting with high energy theory. As it turns out, the ${\bf16}\oplus {\bf 16'}$ Majorana-Weyl (semi)spinors of $\mathfrak{so}_{1,9}$ contain the same physical content as the complex Weyl (semi)spinors ${\bf 16}\oplus \overline{{\bf 16}}$ of $\mathfrak{so}_{10}$. While $\mathfrak{so}_{10}$ spinors separate dofs into left and right chiralities, $\mathfrak{so}_{1,9}$ spinors separates dofs into particles and antiparticles. Adding multiple $\mathfrak{so}_{1,1}$ lightcones allows for multiple mass generations, and including additional off-shell dofs can be thought about as a fourth lightcone, giving $\mathfrak{so}_{4,12}$. As we will show, chiral $\mathfrak{so}_{10}$ spinors can be found from $\mathfrak{e}_{8(-24)}$.

The notion of $\mathfrak{so}_{1,9}$ charge space will be pursued in more detail in future work, but the primary goal of this work is to establish the high energy theory inside $\mathfrak{e}_{8(-24)}$. We briefly note that in addition to high energy GUTs, $\mathfrak{e}_{8(-24)}$ also contains a dual Lorentz symmetry, which was sought after in an attempt to understand the origins of the double copy~\cite{Bern:2008qj} and the low-energy nonsupersymmetric field theory limit of the KLT relations in string theory~\cite{Kawai:1985xq}. It is quite curious to find the signature of the critical spacetime dimensions of superstring theory; however, we should stress that this work does not look to find spacetime inside $\mathfrak{so}_{1,9}$. Isolating $\mathfrak{so}_6$ for Pati-Salam GUT and the strong force would lead to a commuting $\mathfrak{so}_{1,3}$. This seems to be an internal symmetry that mirrors spacetime and allows for a dual Lorentz symmetry different than the one found in pure gravity~\cite{Cheung:2016say,Cheung:2017kzx}.

Similar to how there are three distinct $\mathfrak{e}_7$ subalgebras in $\mathfrak{e}_8$, there are also three distinct $\mathfrak{so}_{10}$'s inside $\mathfrak{e}_6$ and three $\mathfrak{so}_9$'s inside $\mathfrak{f}_4$. These three distinct $\mathfrak{so}_{10}$'s can be found inside $\mathfrak{e}_{8(-24)}$, which simultaneously isolates three distict $\mathfrak{so}_{3,1}$'s, which fit inside $\mathfrak{so}_{3,3}$. In addition to $\mathfrak{so}_{3,1}$ spacetime, $\mathfrak{so}_{4,2}$ is found, which can either be used as a conformal symmetry in $3+1$ spacetime dimensions, or allow for the introduction of $AdS_5 =SO(4,2)/SO(4,1)$ for a single generation. We propose a three-time generalization of $AdS_5$, which is $SO(4,4)/SO(4,3)$ and yields $dS_4 \times S^2$ with $S^2$ for mass/flavor oscillations, generalizing what was found in Ref.~\cite{Lisi:2015oja}. Isolating $\mathfrak{so}_{4,4}$ in $\mathfrak{so}_{4,12}$ leaves behind a commuting $\mathfrak{so}_8$. However, in order to connect with $\mathfrak{so}_{10}$ in $\mathfrak{e}_{8(-24)}$, one must isolate a single generation.

\subsection{From $\mathfrak{e}_{8(-24)}$ to $SO(10)$ GUT with spacetime: a threefold way }
\label{so10}

Using $\mathfrak{so}_{10}$ for GUT is the most popular model, as it unifies the bosons into a single gauge group and a single generation of fermions into a single chiral spinor ${\bf 16}$. Three related ways to break to $\mathfrak{so}_{10}$ and include $\mathfrak{so}_{3,1}$ for spacetime,
\begin{eqnarray}
&& {\bf I} : \mathfrak{so}_{3,11}\oplus \mathfrak{so}_{1,1} \nonumber \\
&\nearrow&\qquad\qquad\qquad \searrow \nonumber \\
\mathfrak{e}_{8(-24)} \rightarrow \mathfrak{so}_{4,12} &\rightarrow& {\bf II} : \mathfrak{so}_{3,1}\oplus \mathfrak{so}_{1,11} \rightarrow \mathfrak{so}_{3,1}\oplus\mathfrak{so}_{10}\oplus\mathfrak{so}_{1,1}. \\
&\searrow& \qquad\qquad\qquad \nearrow \nonumber \\
&& \,\, {\bf III} : \mathfrak{so}_{4,2}\oplus\mathfrak{so}_{10} \nonumber
\label{terzo}
\end{eqnarray}
In generalizing Nesti and Percacci's $\mathfrak{so}_{3,11}$ model~\cite{Nesti:2009kk,Percacci:2009ij} to $\mathfrak{e}_{8(-24)}$ with $\mathfrak{so}_{4,12}$, the following path of symmetry breaking is taken:
\begin{eqnarray}
{\bf I} : \mathfrak{e}_{8(-24)} &\rightarrow& \mathfrak{so}_{4,12} \rightarrow \mathfrak{so}_{3,11} \oplus \mathfrak{so}_{1,1} \rightarrow \mathfrak{so}_{3,1} \oplus \mathfrak{so}_{10} \oplus \mathfrak{so}_{1,1}, \nonumber \\
{\bf 248} &=& {\bf 120} \oplus {\bf 128} = {\bf 91}_0 \oplus {\bf 1}_0 \oplus {\bf 14}_2 \oplus {\bf 14}_{-2} \oplus {\bf 64}_1 \oplus {\bf 64}'_{-1}\\
&=& ({\bf 3},{\bf 1},{\bf 1})_0 \oplus ({\bf 1},{\bf 3},{\bf 1})_0 \oplus ({\bf 1},{\bf 1},{\bf 1})_0 \oplus ({\bf 2},{\bf 2},{\bf 10})_0 \oplus ({\bf 1},{\bf 1},{\bf 45})_0 \nonumber \\
&& \oplus ({\bf 2},{\bf 2},{\bf 1})_2 \oplus ({\bf 1},{\bf 1},{\bf 10})_2 \oplus ({\bf 2},{\bf 2},{\bf 1})_{-2} \oplus ({\bf 1},{\bf 1},{\bf 10})_{-2} \nonumber \\
&& \oplus ({\bf 2},{\bf 1},{\bf 16})_1 \oplus ({\bf 1},{\bf 2},\overline{{\bf 16}})_1 \oplus ({\bf 2},{\bf 1},\overline{{\bf 16}})_{-1} \oplus ({\bf 1},{\bf 2},{\bf 16})_{-1}. \nonumber
\label{primo}
\end{eqnarray}
where the breaking to $\mathfrak{so}_{11,3}\oplus \mathfrak{so}_{1,1}$ yields the extended Poincar\'{e} five-grading of $\mathfrak{e}_{8(-24)}$ mentioned in Ref.~\cite{2014Douglas}.
While typical $\mathfrak{so}_{10}$ GUT analysis refers to on-shell dofs via ${\bf 16}\oplus\overline{{\bf 16}}$, including $\mathfrak{so}_{3,1}$ allows for the off-shell dofs to be accounted for, introducing $({\bf 2},{\bf 1})$ for left chiralities and $({\bf 1},{\bf 2})$ for right chiralities. The $\mathfrak{so}_{1,1}$ weight here is $+1$ for SM fermions and antifermions, while $-1$ gives the mirror fermions. Simply starting with $\mathfrak{so}_{4,12}$ and its spinors only gives one generation and a mirror fermion. However, the algebraic structure here is richer, as $\mathfrak{e}_8$ contains three $\mathfrak{e}_7$'s, which have a fourth ``generation'' shared amongst the others (as it can be seen from the magic star projection of $\mathfrak{e}_8$; see Fig.~\eqref{e8magic}), giving three on-shell generations.

A Higgs candidate is found in $({\bf 1},{\bf 1},{\bf 10})_2$ in ~\eqref{primo}, which was not found in Nesti and Percacci's model~\cite{Nesti:2009kk,Percacci:2009ij}. Various bosonic vectors are also found. These additional dofs will be explored in subsequent work and are outside the scope of this paper.

Next, as another possible option, the spacetime can be isolated from the beginning and shown to give the same result when breaking $\mathfrak{so}_{1,11}$,
\begin{eqnarray}
{\bf II} : \mathfrak{e}_{8(-24)} &\rightarrow& \mathfrak{so}_{4,12} \rightarrow \mathfrak{so}_{3,1}\oplus\mathfrak{so}_{1,11} \rightarrow \mathfrak{so}_{3,1} \oplus \mathfrak{so}_{10}\oplus \mathfrak{so}_{1,1} \nonumber \\
{\bf 248} &=& {\bf 120} \oplus {\bf 128} \\
&=& ({\bf 3},{\bf 1},{\bf 1}) \oplus ({\bf 1},{\bf 3},{\bf 1}) \oplus ({\bf 1},{\bf 1},{\bf 66}) \oplus ({\bf 2},{\bf 2},{\bf 12}) \oplus ({\bf 2},{\bf 1},{\bf 32}) \oplus ({\bf 1},{\bf 2},\overline{{\bf 32}}) \nonumber \\
&=& ({\bf 3},{\bf 1},{\bf 1})_{0} \oplus ({\bf 1},{\bf 3},{\bf 1})_{0} \oplus ({\bf 1},{\bf 1},{\bf 45})_{0} \oplus ({\bf 1},{\bf 1},{\bf 1})_{0} \nonumber \\
&& \oplus ({\bf 1},{\bf 1},{\bf 10})_{2} \oplus ({\bf 1},{\bf 1},{\bf 10})_{-2} \oplus ({\bf 2},{\bf 2},{\bf 10})_{0} \oplus ({\bf 2},{\bf 2},{\bf 1})_{2} \oplus ({\bf 2},{\bf 2},{\bf 1})_{-2} \nonumber \\
&& \oplus ({\bf 2},{\bf 1},{\bf 16})_{1} \oplus ({\bf 2},{\bf 1},\overline{{\bf 16}})_{-1} \oplus ({\bf 1},{\bf 2},\overline{{\bf 16}})_{1} \oplus ({\bf 1},{\bf 2},{\bf 16})_{-1}. \nonumber
\end{eqnarray}
Removing $\mathfrak{so}_{3,1}$ isolates $\mathfrak{so}_{1,11}$, which happens to be the spacetime signature of F-theory. The difference here is that $\mathfrak{so}_{1,11}$ is used for a type of Lorentzian charge space, rather than spacetime. Breaking off the charge space lightcone $\mathfrak{so}_{1,1}$ isolates $\mathfrak{so}_{10}$ and splits the $({\bf 2},{\bf 1},{\bf 32})$ of $\mathfrak{so}_{1,11}$ into a left-handed ${\bf 16}$ spinor of $\mathfrak{so}_{10}$ with its mirror $\overline{{\bf 16}}$, given by $ ({\bf 2},{\bf 1},{\bf 16})_{1}$ and $({\bf 2},{\bf 1},\overline{{\bf 16}})_{-1}$, respectively.

Finally, as the third possibility indicated in ~\eqref{terzo}, one may also immediately isolate $\mathfrak{so}_{10}$ to give $\mathfrak{so}_{4,2}$,
\begin{eqnarray}
{\bf III} : \mathfrak{e}_{8(-24)} &\rightarrow& \mathfrak{so}_{4,12} \rightarrow \mathfrak{so}_{4,2} \oplus \mathfrak{so}_{10} \rightarrow \mathfrak{so}_{3,1}\oplus \mathfrak{so}_{10} \oplus \mathfrak{so}_{1,1} \nonumber \\
{\bf 248} &=& {\bf 120} \oplus {\bf 128} = ({\bf 15},{\bf 1}) \oplus ({\bf 1},{\bf 45}) \oplus ({\bf 6},{\bf 10}) \oplus ({\bf 4 },{\bf 16}) \oplus (\overline{{\bf 4}},\overline{{\bf 16}}) \\
&=& ({\bf 3},{\bf 1},{\bf 1})_{0} \oplus ({\bf 1},{\bf 3},{\bf 1})_{0} \oplus ({\bf 1},{\bf 1},{\bf 1})_{0} \oplus ({\bf 2},{\bf 2},{\bf 1})_{2} \oplus ({\bf 2},{\bf 2},{\bf 1})_{-2} \nonumber \\
&& \oplus ({\bf1},{\bf1},{\bf 45})_{0} \oplus ({\bf 2},{\bf 2},{\bf 1})_{0} \oplus ({\bf 1},{\bf 1},{\bf 1})_{2} \oplus ({\bf 1},{\bf 1},{\bf 1})_{-2} \nonumber \\
&& \oplus ({\bf 2},{\bf 1},{\bf 16})_{1} \oplus ({\bf 1},{\bf 2},{\bf 16})_{-1} \oplus ({\bf 1},{\bf 2},\overline{{\bf 16}})_{1} \oplus ({\bf 2},{\bf 1},\overline{{\bf 16}})_{-1}. \nonumber
\end{eqnarray}
This approach gives $\mathfrak{so}_{4,2}$, which may be either the conformal symmetry of $\mathfrak{so}_{3,1}$ or the algebra of isometries of $AdS_5$ (which can be broken to $dS_4$~\cite{Chu:2016uwi}). While $\mathfrak{so}_{3,1}$ is useful for 4D physics, $dS_4$ is applicable for an expanding universe with a positive cosmological constant. While typical GUT refers to the on-shell fermionic dofs only, we find that including $\mathfrak{so}_{3,1}\sim \mathfrak{sl}_{2,\mathbb{C}}$ allows for the identification of off-shell fermions, such as $({\bf2},{\bf1},{\bf16})_1$.

Since $\mathfrak{so}_{10}$ GUT can be embedded inside $\mathfrak{e}_{6(-78)}$ and $\mathfrak{e}_{6(-14)}$, it is also possible to break $\mathfrak{e}_{8(-24)}$ to either real form of $\mathfrak{e}_6$ \cite{dGM} and obtain $\mathfrak{so}_{10}$,
\begin{eqnarray}
\mathfrak{e}_{8(-24)} &\rightarrow& \mathfrak{e}_{6(-78)}\oplus \mathfrak{su}_{2,1}\nonumber \\
\downarrow&& \mbox{ }\mbox{ }\mbox{ }\mbox{ }\mbox{ }\mbox{ }\mbox{ }\mbox{ }\mbox{ }\downarrow \nonumber \\
\mathfrak{e}_{6(-14)}\oplus \mathfrak{su}_{2,1} &\rightarrow& \mathfrak{so}_{10}\oplus \mathfrak{su}_{2,1} \oplus \mathfrak{u}_1,
\end{eqnarray}
However, this does not allow for the isolation of $\mathfrak{so}_{3,1}$ spacetime. Nevertheless, $\mathfrak{e}_6$ may isolate three on-shell generations from the ``fourth'' additional off-shell generation,
\begin{eqnarray}
\mathfrak{e}_{8(-24)} &\rightarrow& \mathfrak{e}_{6(-78)}\oplus\mathfrak{su}_{2,1} \rightarrow \mathfrak{so}_{10} \oplus \mathfrak{su}_{2,1} \oplus \mathfrak{u}_1 \nonumber \\
{\bf 248} &=& ({\bf 78},{\bf 1}) \oplus ({\bf 1},{\bf 8}) \oplus ({\bf 27},{\bf 3}) \oplus (\overline{{\bf 27}},\overline{{\bf 3}}) \\
&=& ({\bf 45},{\bf 1})_0 \oplus ({\bf 1},{\bf 8})_0 \oplus ({\bf 1},{\bf 1})_0 \oplus ({\bf 16},{\bf 1})_{-3} \oplus (\overline{{\bf 16}},{\bf 1})_3 \nonumber \\
&& \oplus ({\bf 16},{\bf 3})_1 \oplus ({\bf 10},{\bf 3})_{-2} \oplus ({\bf 1},{\bf 3})_4 \oplus (\overline{{\bf 16}},\overline{{\bf 3}})_{-1} \oplus ({\bf10},\overline{{\bf 3}})_2 \oplus ({\bf 1},\overline{{\bf 3}})_{-4}, \nonumber
\end{eqnarray}
as the ${\bf 4}$ of $\mathfrak{so}_{4,2}$ gets separated to ${\bf 3}\oplus{\bf 1}$ of $\mathfrak{su}_{2,1}$. 
It appears that $\mathfrak{e}_6$ doesn't directly refer to mirror fermions, while $\mathfrak{so}_{10}$ does. It's also worth noting that this approach to $\mathfrak{e}_{6(-78)}$ is dissimilar to $E_6$ GUT, as the typical $E_6$ GUT introduces additional fermions into the ${\bf 27}$ of $E_6$, while this approach only assigns fermions to the ${\bf 16}$ of the Peirce decomposition of ${\bf 27}$. Furthermore, the interpretation of ${\bf 27}$ as an exceptional Jordan algebra over $\mathbb{R}$ only occurs with $\mathfrak{e}_{6(-26)}$ and $\mathfrak{e}_{6(6)}$, which respectively contains $\mathfrak{so}_{1,9}\oplus \mathfrak{so}_{1,1}$ and $\mathfrak{so}_{5,5}\oplus\mathfrak{so}_{1,1}$ (reduced structure algebras of $\mathbb{R}\oplus \mathfrak{J}_{2}\left( \mathbb{O}\right) $ and of $\mathbb{R}\oplus \mathfrak{J}_{2}\left( \mathbb{O}_{s}\right) $, respectively) as a subalgebra, rather than $\mathfrak{so}_{10}\oplus\mathfrak{u}_1$. The complete comparison of $\mathfrak{so}_{10}$ and $\mathfrak{so}_{1,9}$ is saved for future work, and we will not develop a full-fledged $E_6$ GUT model here, as there are many possibilities to consider, thus deserving a separate treatment.

\subsection{From $\mathfrak{e}_{8(-24)}$ to Pati-Salam GUT with spacetime: a twofold way }\label{PS}

Since it is already understood that $\mathfrak{su}_4\oplus\mathfrak{su}_2\oplus\mathfrak{su}_2$ GUT can be found from $\mathfrak{so}_{10}$, obtaining Pati-Salam GUT from $\mathfrak{e}_8$ is trivial, since Section~\eqref{so10} found $\mathfrak{so}_{10}$ from $\mathfrak{e}_8$. Now, we focus on including reference to spacetime to explicitly confirm chiralities. 

To start, we break $\mathfrak{e}_{8(-24)}$ through $\mathfrak{so}_{10}$ to Pati-Salam with $\mathfrak{so}_{4,2}$ and then to $\mathfrak{so}_{3,1}$ to ensure the appropriate chiralities and to confirm, as pointed out above, that the $\mathfrak{so}_{1,1}$ weight refers to nonmirror or mirror fermions. Breaking $\mathfrak{e}_{8(-24)}$ to $\mathfrak{so}_{10}$ and $\mathfrak{su}_{4}\oplus\mathfrak{su}_2\oplus\mathfrak{su}_2$ gives
\begin{eqnarray}
\mathfrak{e}_{8(-24)} &\rightarrow& \mathfrak{so}_{4,12} \rightarrow \mathfrak{so}_{4,2} \oplus \mathfrak{so}_{10} \rightarrow \mathfrak{so}_{4,2} \oplus \mathfrak{su}_{4} \oplus \mathfrak{su}_2\oplus\mathfrak{su}_2 \rightarrow \mathfrak{so}_{3,1} \oplus \mathfrak{su}_4 \oplus \mathfrak{su}_2 \oplus \mathfrak{su}_2 \oplus \mathfrak{so}_{1,1}\nonumber \\
{\bf 248} &=& {\bf 120} \oplus {\bf 128} = ({\bf 15},{\bf 1}) \oplus ({\bf 1},{\bf 45}) \oplus ({\bf 6},{\bf 10}) \oplus ({\bf 4},{\bf 16}) \oplus (\overline{{\bf 4}},\overline{{\bf 16}}) \\
&=& ({\bf 15},{\bf 1},{\bf 1},{\bf 1}) \oplus ({\bf 1},{\bf 15}, {\bf 1}, {\bf 1}) \oplus ({\bf 1},{\bf 1},{\bf 3},{\bf1}) \oplus ({\bf 1},{\bf 1},{\bf 1},{\bf 3}) \oplus ({\bf 1},{\bf 6},{\bf 2},{\bf 2}) \nonumber \\
&& \oplus ({\bf 6},{\bf 6},{\bf 1},{\bf 1}) \oplus ({\bf 6},{\bf 1},{\bf 2},{\bf 2}) \oplus ({\bf 4},{\bf 4},{\bf 2},{\bf 1}) \oplus ({\bf
4},\overline{{\bf 4}},{\bf 1},{\bf 2}) \oplus (\overline{{\bf 4}},{\bf 4},{\bf 1},{\bf 2}) \oplus (\overline{{\bf 4}},\overline{{\bf 4}},{\bf 2},{\bf 1}) \nonumber \\
&=& ({\bf 3},{\bf 1},{\bf 1},{\bf 1},{\bf 1})_{0} \oplus ({\bf 1},{\bf 3},{\bf 1},{\bf 1},{\bf 1})_{0} \oplus ({\bf 1},{\bf 1},{\bf 1},{\bf 1},{\bf 1})_{0} \oplus ({\bf 2},{\bf 2},{\bf 1},{\bf 1},{\bf 1})_{2} \nonumber \\
&& \oplus ({\bf 2},{\bf 2},{\bf 1},{\bf 1},{\bf 1})_{-2} \oplus ({\bf 1},{\bf 1},{\bf 15},{\bf 1},{\bf 1})_{0} \oplus ({\bf 1},{\bf 1},{\bf 1},{\bf 3},{\bf 1})_{0} \oplus ({\bf 1},{\bf 1},{\bf 1},{\bf 1},{\bf 3})_{0} \nonumber \\
&& \oplus ({\bf 1},{\bf 1},{\bf 6},{\bf 2},{\bf 2})_{0} \oplus ({\bf 2},{\bf 2},{\bf 6},{\bf 1},{\bf 1})_{0} \oplus ({\bf 1},{\bf 1},{\bf 6},{\bf 1},{\bf 1})_{2} \oplus ({\bf 1},{\bf 1},{\bf 6},{\bf 1},{\bf 1})_{-2} \nonumber \\
&& \oplus ({\bf 2},{\bf 2},{\bf 1},{\bf 2},{\bf 2})_{0} \oplus ({\bf 1},{\bf 1},{\bf 1},{\bf 2},{\bf 2})_{2} \oplus ({\bf 1},{\bf 1},{\bf 1},{\bf 2},{\bf 2})_{-2} \nonumber \\
&& \oplus ({\bf 2},{\bf 1},{\bf 4},{\bf 2},{\bf 1})_{1} \oplus ({\bf 1},{\bf 2},{\bf 4},{\bf 2},{\bf 1})_{-1} \oplus ({\bf 2},{\bf 1},\overline{{\bf 4}},{\bf 1},{\bf 2})_{1} \oplus ({\bf 1},{\bf 2},\overline{{\bf 4}},{\bf 1},{\bf 2})_{-1} \nonumber \\
&& \oplus ({\bf 1},{\bf 2},{\bf 4},{\bf 1},{\bf 2})_{1} \oplus ({\bf 2},{\bf 1},{\bf 4},{\bf 1},{\bf 2})_{-1} \oplus ({\bf 1},{\bf 2},\overline{{\bf 4}},{\bf 2},{\bf 1})_{1} \oplus ({\bf 2},{\bf 1},\overline{{\bf 4}},{\bf 2},{\bf 1})_{-1}. \nonumber
\end{eqnarray}
The fermions with $\mathfrak{so}_{1,1}$ weight $1$ correspond to one generation, while -1 weights correspond to mirror fermions.

Next, bypassing $\mathfrak{so}_{10}$,a different approach to get Pati-Salam GUT that bypasses $\mathfrak{so}_{10}$ is discussed. The unification of the three forces via GUT seems computationally motivated by the almost unification of the coupling constants of the strong and electroweak forces~\cite{Langacker:1980js}. This with the combination of difficulty of treating general relativity as a quantum field theory tends to unify the strong force with the electroweak force \textit{before} gravity. However, treating gravity as a gauge theory may help. In particular, the frame field can be used for a Higgs-like mechanism~\cite{Chisholm1992,Krasnov:2011hi} for breaking from higher to lower dimensions~\cite{Das:2018umm}. Also, the dilaton relates to conformal symmetry breaking and has been proposed as a Higgs candidate~\cite{Goldberger:2008zz}. Furthermore, the electroweak Higgs boson provides mass, which is a charge of gravity. 

Since it appears that trivially combining the strong force with the electroweak force under a single gauge group leads to proton decay, we demonstrate a way to unify spacetime with the electroweak force. Starting from $\mathfrak{e}_{8(-24)}$, a new path to break to Pati-Salam GUT is found that bypasses $SO(10)$ GUT:
\begin{eqnarray}
\mathfrak{e}_{8(-24)} &\rightarrow& \mathfrak{so}_{4,12} \rightarrow \mathfrak{so}_{3,5} \oplus \mathfrak{so}_{1,7} \rightarrow \mathfrak{so}_{3,1} \oplus \mathfrak{su}_2 \oplus \mathfrak{su}_2 \oplus \mathfrak{so}_{1,7} \rightarrow \mathfrak{so}_{3,1} \oplus \mathfrak{su}_2 \oplus \mathfrak{su}_2 \oplus \mathfrak{su}_4 \oplus \mathfrak{so}_{1,1} \nonumber \\
{\bf 248} &=& {\bf 120} \oplus {\bf 128} = ({\bf 28},{\bf 1}) \oplus ({\bf 1},{\bf 28}) \oplus ({\bf 8}_v,{\bf8}_v) \oplus ({\bf 8}_s,{\bf 8}_c) \oplus ({\bf 8}_c,{\bf 8}_s) \\
&=& ({\bf 3},{\bf1},{\bf 1},{\bf 1},{\bf 1}) \oplus ({\bf 1},{\bf 3},{\bf 1},{\bf 1},{\bf 1}) \oplus ({\bf 1},{\bf 1},{\bf 3},{\bf 1},{\bf 1}) \oplus ({\bf 1},{\bf 1},{\bf 1},{\bf 3},{\bf 1}) \nonumber \\
&& \oplus ({\bf 2},{\bf 2},{\bf 2},{\bf 2},{\bf 1}) \oplus ({\bf 1},{\bf 1},{\bf 1},{\bf 1},{\bf 28}) \oplus ({\bf 2},{\bf 2},{\bf 1},{\bf 1},{\bf 8}_v) \oplus ({\bf 1},{\bf 1},{\bf 2},{\bf 2},{\bf 8}_v) \nonumber \\
&& \oplus ({\bf 2},{\bf 1},{\bf 2},{\bf 1},{\bf 8}_c) \oplus ({\bf 1},{\bf 2},{\bf 1},{\bf 2},{\bf 8}_c) \oplus ({\bf 2},{\bf 1},{\bf 1},{\bf 2},{\bf 8}_s) \oplus ({\bf1},{\bf 2},{\bf 2},{\bf 1},{\bf 8}_s) \nonumber \\
&=& ({\bf 3},{\bf1},{\bf 1},{\bf 1},{\bf 1})_0 \oplus ({\bf 1},{\bf 3},{\bf 1},{\bf 1},{\bf 1})_0 \oplus ({\bf 1},{\bf 1},{\bf 3},{\bf 1},{\bf 1})_0 \oplus ({\bf 1},{\bf 1},{\bf 1},{\bf 3},{\bf 1})_0 \nonumber \\
&& \oplus ({\bf 2},{\bf 2},{\bf 2},{\bf 2},{\bf 1})_0 \oplus ({\bf 1},{\bf 1},{\bf 1},{\bf 1},{\bf 15})_0 \oplus ({\bf 1},{\bf 1},{\bf 1},{\bf 1},{\bf 1})_0 \oplus ({\bf 1},{\bf 1},{\bf 1},{\bf 1},{\bf 6})_2 \nonumber \\
&& \oplus ({\bf 1},{\bf 1},{\bf 1},{\bf 1},{\bf 6})_{-2} \oplus ({\bf 2},{\bf 2},{\bf 1},{\bf 1},{\bf 6})_{0} \oplus ({\bf 2},{\bf 2},{\bf 1},{\bf 1},{\bf 1})_{2} \oplus ({\bf 2},{\bf 2},{\bf 1},{\bf 1},{\bf 1})_{-2} \nonumber \\
&& \oplus ({\bf 1},{\bf 1},{\bf 2},{\bf 2},{\bf 6})_{0} \oplus ({\bf 1},{\bf 1},{\bf 2},{\bf 2},{\bf 1})_{2} \oplus ({\bf 1},{\bf 1},{\bf 2},{\bf 2},{\bf 1})_{-2} \nonumber \\
&&\oplus ({\bf 2},{\bf 1},{\bf 2},{\bf 1},\overline{{\bf 4}})_1 \oplus ({\bf 2},{\bf 1},{\bf 2},{\bf 1},{\bf 4})_{-1} \oplus ({\bf 1},{\bf 2},{\bf 1},{\bf 2},\overline{{\bf 4}})_1  \oplus ({\bf 1},{\bf 2},{\bf 1},{\bf 2},{\bf 4})_{-1} \nonumber \\
&& \oplus ({\bf 2},{\bf 1},{\bf 1},{\bf 2},{\bf 4})_1 \oplus ({\bf 2},{\bf 1},{\bf 1},{\bf 2},\overline{{\bf 4}})_{-1} \oplus ({\bf1},{\bf 2},{\bf 2},{\bf 1},{\bf 4})_1 \oplus ({\bf1},{\bf 2},{\bf 2},{\bf 1},\overline{{\bf 4}})_{-1}. \nonumber
\end{eqnarray}
As shown above, this route avoids $\mathfrak{so}_{10}$, which is known to give proton decay. It turns out that the mirror fermions have weight $+1$ this time. While further work is needed to systematically determine if this path avoids proton decay, this approach presents a candidate, although there are nontrivial alternatives with $\mathfrak{su}_5$~\cite{Arnowitt1993,Altarelli2001}, as well.

%

\subsection{From $\mathfrak{e}_{8(-24)}$ to $SU(5)$ GUT with spacetime: a threefold way }\label{Sec3}

Next, we demonstrate that there are three different ways to symmetry break $\mathfrak{e}_{8(-24)}$ and obtain $\mathfrak{su}_5$ GUT with spacetime. The most straightforward one breaks from $\mathfrak{so}_{10}$, since Section~\eqref{so10} found $\mathfrak{so}_{10}$ inside $\mathfrak{e}_{8(-24)}$. Since $\mathfrak{e}_{8(-24)}$ has $\mathfrak{su}_{2,7}$ as a maximal and non-symmetric subalgebra \cite{dGM}, $\mathfrak{su}_5$ with spacetime can be recovered, as $\mathfrak{su}_{2,2}\sim \mathfrak{so}_{4,2}$, by breaking $\mathfrak{su}_{2,7}$ to $\mathfrak{su}_{2,2}\oplus \mathfrak{su}_{5}$. The $\mathfrak{su}_7$ allows for the cohomology description of the fermions~\cite{Dijkgraaf:1996hk,Rios:2019rfc}. Additionally, $\mathfrak{su}_{2,3}\oplus\mathfrak{su}_5$ is also a subalgebra of $\mathfrak{e}_{8(-24)}$ \cite{dGM}. This provides at least three distinct ways to break from $\mathfrak{e}_{8(-24)}$ to $\mathfrak{su}_{5}\oplus \mathfrak{so}_{4,2}$,
\begin{equation}
\begin{array}{cclccc}
& \nearrow &{\bf 1}:& \mathfrak{so}_{4,12}\rightarrow \mathfrak{so}_{10}\oplus \mathfrak{su}_{2,2} & \searrow &  \\
\mathfrak{e}_{8(-24)} & \rightarrow & {\bf 2}:& \mathfrak{su}_{5}\oplus \mathfrak{su}_{2,3} &
\rightarrow & \mathfrak{su}_{5}\oplus \mathfrak{so}_{4,2}\oplus
\mathfrak{u}_{1} \\
& \searrow & {\bf 3}:& \mathfrak{su}_{2,7} & \nearrow &
\end{array}
\end{equation}

Breaking from $\mathfrak{e}_{8(-24)}$ through $\mathfrak{so}_{10}$ and to $\mathfrak{su}_5$ gives
\begin{eqnarray}
{\bf1}:\, \mathfrak{e}_{8(-24)} &\rightarrow& \mathfrak{so}_{4,12} \rightarrow \mathfrak{so}_{10}\oplus \mathfrak{so}_{4,2} \rightarrow \mathfrak{su}_5 \oplus \mathfrak{so}_{4,2}\oplus\mathfrak{u}_1, \\
{\bf 248} &=& {\bf 120} \oplus {\bf 128} = ({\bf 45},{\bf 1}) \oplus ({\bf 1},{\bf 15}) \oplus ({\bf 10},{\bf 6}) \oplus ({\bf 16},{\bf 4}) \oplus (\overline{{\bf 16}},\overline{{\bf 4}}) \nonumber \\
&=& ({\bf 24},{\bf 1})_{0} \oplus ({\bf 1},{\bf 1})_0 \oplus ({\bf 10},{\bf 1})_4 \oplus (\overline{{\bf 10}},{\bf 1})_{-4} \oplus ({\bf 1},{\bf 15})_0 \oplus ({\bf 5},{\bf 6})_2 \oplus (\overline{{\bf 5}},{\bf 6})_{-2} \nonumber \\
&& \oplus ({\bf 10},{\bf 4})_{-1} \oplus (\overline{{\bf 5}},{\bf 4})_3 \oplus ({\bf1},{\bf 4})_{-5} \oplus (\overline{{\bf 10}},\overline{{4}})_1 \oplus ({\bf 5},\overline{{\bf 4}})_{-3} \oplus ({\bf 1},\overline{{\bf 4}})_5. \nonumber
\end{eqnarray}
As shown in Eq.~\eqref{SU5break}, $({\bf 10},{\bf 4})_{-1}$ of $\mathfrak{su}_5\oplus\mathfrak{so}_{4,2}$ contains left-handed quarks, anti-up quarks, and the positron, while $(\overline{{\bf 5}},{\bf 4})_{3}$ contains left-handed anti-down quarks and leptons (of a single generation with their mirror fermions). As mentioned above, once $\mathfrak{su}_{2,2}\sim\mathfrak{so}_{4,2}$ is obtained, this may be used as the isometry of $AdS_5$ and broken to $dS_4$ or simply broken to $\mathfrak{so}_{3,1}\oplus\mathfrak{so}_{1,1}$.


Let us start and consider four different paths from $\mathfrak{su}_5\oplus\mathfrak{su}_{3,2}$,%
\begin{equation}
\mathbf{2}:\mathfrak{e}_{8(-24)}\rightarrow \left\{
\begin{array}{l}
\left( \mathfrak{su}_{5}\oplus\mathfrak{su}_{3,2}\right) _{I}\rightarrow
\mathfrak{su}_{5}\oplus\mathfrak{su}_{2,2}\oplus\mathfrak{u}_{1}\rightarrow \left\{
\begin{array}{l}
\mathfrak{su}_{5}\oplus\mathfrak{so}_{3,1,a}\oplus\mathfrak{u}_{1}\oplus\mathfrak{so}_{1,1}, \\
\\
\mathfrak{su}_{5}\oplus\mathfrak{so}_{3,1,b}\oplus\mathfrak{u}_{1}\oplus\mathfrak{so}_{1,1},%
\end{array}%
\right.  \\
\\
\left( \mathfrak{su}_{5}\oplus\mathfrak{su}_{3,2}\right) _{II}\rightarrow
\mathfrak{su}_{5}\oplus\mathfrak{su}_{2,2}\oplus\mathfrak{u}_{1}\rightarrow\left\{
\begin{array}{l}
\mathfrak{su}_{5}\oplus\mathfrak{so}_{3,1,a}\oplus\mathfrak{u}_{1}\oplus\mathfrak{so}_{1,1}, \\
\\
\mathfrak{su}_{5}\oplus\mathfrak{so}_{3,1,b}\oplus\mathfrak{u}_{1}\oplus\mathfrak{so}_{1,1},%
\end{array}%
\right.  \\%
\end{array}%
\right.
\end{equation}%
where%
\begin{eqnarray}
I &:&\left\{
\begin{array}{l}
\mathfrak{e}_{8(-24)}\rightarrow \mathfrak{su}_{5}\oplus\mathfrak{su}_{3,2}, \\
\mathbf{248}=\left( \mathbf{24},\mathbf{1}\right) +\left( \mathbf{1},\mathbf{%
24}\right) +\left( \mathbf{10},\mathbf{5}\right) +\left( \overline{\mathbf{10%
}},\overline{\mathbf{5}}\right) +\left( \mathbf{5},\overline{\mathbf{10}}%
\right) +\left( \overline{\mathbf{5}},\mathbf{10}\right),%
\label{primero}
\end{array}%
\right.  \\
II &:&\left\{
\begin{array}{l}
\mathfrak{e}_{8(-24)}\rightarrow \mathfrak{su}_{5}\oplus\mathfrak{su}_{3,2}, \\
\mathbf{248}=\left( \mathbf{24},\mathbf{1}\right) +\left( \mathbf{1},\mathbf{%
24}\right) +\left( \overline{\mathbf{10}},\mathbf{5}%
\right) +\left( \mathbf{10},\overline{\mathbf{5}}\right) +\left( \mathbf{5},\mathbf{10}\right) +\left( \overline{\mathbf{5}%
},\overline{\mathbf{10}}\right) .%
\end{array}%
\right.
\label{segundo}
\end{eqnarray}%
It is evident that \eqref{primero} and \eqref{segundo} are related by an exchange $\mathfrak{su}_{5}\leftrightarrow \mathfrak{su}_{3,2}$. In $\mathbb{C}$, the two $\mathfrak{a}_{4}\sim \mathfrak{su}_5$ in $\mathfrak{e}_{8}$ are not equivalent, thus their embedding
in $\mathfrak{e}_{8}$ is not symmetric under the exchange of them\footnote{The two conjugacy classes of subalgebras are related by an internal automorphisms of $\mathfrak{e}_{8(-24)}$, corresponding to conjugation of modules within the second summand in $\mathfrak{a}_{4}\oplus \mathfrak{a}_{4}\subset \mathfrak{e}_{8}$.}.

Moreover, there are also two conjugacy classes of $\mathfrak{so}_{3,1}\oplus \mathbb{R}\subset \mathfrak{su}_{2,2}$, related by the conjugation (flip) of the weights of $\mathfrak{so}_{1,1}$:
\begin{eqnarray}
a &:&\left\{
\begin{array}{l}
\mathfrak{su}_{2,2}\rightarrow \mathfrak{so}_{3,1}\oplus\mathfrak{so}_{1,1}, \\
\mathbf{4}=\left( \mathbf{2},\mathbf{1}\right) _{1}\oplus\left( \mathbf{1},%
\mathbf{2}\right) _{-1}, \\
\overline{\mathbf{4}}=\left( \mathbf{2},\mathbf{1}\right) _{-1}\oplus\left(
\mathbf{1},\mathbf{2}\right) _{1}, \\
\mathbf{6}=\left( \mathbf{2},\mathbf{2}\right) _{0}\oplus\left( \mathbf{1},%
\mathbf{1}\right) _{2}\oplus\left( \mathbf{1},\mathbf{1}\right) _{-2},%
\end{array}%
\right.  \label{primmo}\\
b &:&\left\{
\begin{array}{l}
\mathfrak{su}_{2,2}\rightarrow \mathfrak{so}_{3,1}\oplus\mathfrak{so}_{1,1}, \\
\mathbf{4}=\left( \mathbf{2},\mathbf{1}\right) _{-1}\oplus \left( \mathbf{1},%
\mathbf{2}\right) _{1}, \\
\overline{\mathbf{4}}=\left( \mathbf{2},\mathbf{1}\right) _{1}\oplus \left(
\mathbf{1},\mathbf{2}\right) _{-1}, \\
\mathbf{6}=\left( \mathbf{2},\mathbf{2}\right) _{0}\oplus\left( \mathbf{1},%
\mathbf{1}\right) _{2}\oplus\left( \mathbf{1},\mathbf{1}\right) _{-2}.%
\end{array}%
\right. \label{seccondo}
\end{eqnarray}%
The two $\mathfrak{a}_{1}$ inside $\mathfrak{su}_{4}$ are not equivalent, if one consider
the charge with respect to $T_{1}$; thus, the embedding of $\mathfrak{a}_{1}+\mathfrak{a}_{1}+T_{1}
$ into $\mathfrak{a}_{3}$ is not symmetric under the exchange of the two $\mathfrak{a}_{1}$'s.
Note that the exchange $a\leftrightarrow b$ is equivalent to 
flipping the weight associated with $\mathfrak{so}_{1,1}$.

The branching of the $\mathbf{248}$ of $\mathfrak{e}_{8(-24)}$ goes as
follows,
\begin{eqnarray}
\mathbf{2}.I.a &:&\mathbf{248}=\left( \mathbf{24},\mathbf{1}\right) \oplus\left(
\mathbf{1},\mathbf{24}\right) \oplus\left( \mathbf{10},\mathbf{5}\right) \oplus\left(
\overline{\mathbf{10}},\overline{\mathbf{5}}\right) \oplus\left( \mathbf{5},%
\overline{\mathbf{10}}\right) \oplus\left( \overline{\mathbf{5}},\mathbf{10}%
\right)   \nonumber \\
&=&\left( \mathbf{24},\mathbf{1}\right) _{0}\oplus\left( \mathbf{1},\mathbf{15}%
\right) _{0}\oplus\left( \mathbf{1},\overline{\mathbf{4}}\right) _{5}\oplus\left(
\mathbf{1},\mathbf{4}\right) _{-5}\oplus\left( \mathbf{1},\mathbf{1}\right) _{0}
\nonumber \\
&&\oplus\left( \mathbf{10},\mathbf{4}\right) _{-1}\oplus\left( \mathbf{10},\mathbf{1}%
\right) _{4}\oplus\left( \overline{\mathbf{10}},\overline{\mathbf{4}}\right)
_{1}\oplus\left( \overline{\mathbf{10}},\mathbf{1}\right) _{-4}  \nonumber \\
&&\oplus\left( \mathbf{5},\mathbf{6}\right) _{2}\oplus\left( \mathbf{5},\overline{%
\mathbf{4}}\right) _{-3}\oplus\left( \overline{\mathbf{5}},\mathbf{6}\right)
_{-2}\oplus\left( \overline{\mathbf{5}},\mathbf{4}\right) _{3}  \nonumber \\
&=&\left( \mathbf{24},\mathbf{1,1}\right) _{0,0}\oplus\left( \mathbf{1},\mathbf{%
3,1}\right) _{0,0}\oplus\left( \mathbf{1},\mathbf{1,3}\right) _{0,0}\oplus\left(
\mathbf{1},\mathbf{2,2}\right) _{0,2}\oplus\left( \mathbf{1},\mathbf{2,2}\right)
_{0,-2} \nonumber \\
&& \oplus\left( \mathbf{1},\mathbf{1,1}\right) _{0,0} \oplus\left( \mathbf{1},\mathbf{2,1}\right) _{5,-1}\oplus\left( \mathbf{1},\mathbf{%
1,2}\right) _{5,1}\oplus\left( \mathbf{1},\mathbf{2,1}\right)_{-5,1}\oplus\left(
\mathbf{1},\mathbf{1,2}\right) _{-5,-1} \nonumber \\
&& \oplus\left( \mathbf{1},\mathbf{1,1}%
\right) _{0,0} \oplus\left( \mathbf{10},\mathbf{2,1}\right) _{-1,1}\oplus\left( \mathbf{10},\mathbf{%
1,2}\right) _{-1,-1}\oplus\left( \mathbf{10},\mathbf{1,1}\right) _{4,0} \nonumber \\
&& \oplus\left(
\overline{\mathbf{10}},\mathbf{2,1}\right) _{1,-1}\oplus\left( \overline{\mathbf{%
10}},\mathbf{1,2}\right) _{1,1}\oplus\left( \overline{\mathbf{10}},\mathbf{1,1}%
\right) _{-4,0}  \nonumber \\
&&\oplus\left( \mathbf{5},\mathbf{2,2}\right) _{2,0}\oplus\left( \mathbf{5},\mathbf{1,1%
}\right) _{2,2}\oplus\left( \mathbf{5},\mathbf{1,1}\right) _{2,-2}\oplus\left( \mathbf{%
5},\mathbf{2,1}\right) _{-3,-1}\oplus\left( \mathbf{5},\mathbf{1,2}\right) _{-3,1}
\nonumber \\
&&\oplus\left( \overline{\mathbf{5}},\mathbf{2,2}\right) _{-2,0}\oplus\left( \overline{%
\mathbf{5}},\mathbf{1,1}\right) _{-2,2}\oplus\left( \overline{\mathbf{5}},\mathbf{%
1,1}\right) _{-2,-2}\oplus\left( \overline{\mathbf{5}},\mathbf{2,1}\right)
_{3,1}\oplus\left( \overline{\mathbf{5}},\mathbf{1,2}\right) _{3,-1}.\label{1.I.a}
\end{eqnarray}%

\begin{eqnarray}
\mathbf{2}.II.a &:&\mathbf{248}=\left( \mathbf{24},\mathbf{1}\right) \oplus\left(
\mathbf{1},\mathbf{24}\right) \oplus\left( \mathbf{5},\mathbf{10}\right) \oplus\left(
\overline{\mathbf{5}},\overline{\mathbf{10}}\right) \oplus\left( \overline{%
\mathbf{10}},\mathbf{5}\right) \oplus\left( \mathbf{10},\overline{\mathbf{5}}%
\right)   \nonumber \\
&=&\left( \mathbf{24},\mathbf{1}\right) _{0}\oplus\left( \mathbf{1},\mathbf{15}%
\right) _{0}\oplus\left( \mathbf{1},\overline{\mathbf{4}}\right) _{5}\oplus\left(
\mathbf{1},\mathbf{4}\right) _{-5}\oplus\left( \mathbf{1},\mathbf{1}\right) _{0}
\nonumber \\
&&\oplus\left( \mathbf{10},\overline{\mathbf{4}}\right) _{1}\oplus\left( \mathbf{10},%
\mathbf{1}\right) _{-4}\oplus\left( \overline{\mathbf{10}},\mathbf{4}\right)
_{-1}\oplus\left( \overline{\mathbf{10}},\mathbf{1}\right) _{4}  \nonumber \\
&&\oplus\left( \mathbf{5},\mathbf{6}\right) _{-2}\oplus\left( \mathbf{5},\mathbf{4}%
\right) _{3}\oplus\left( \overline{\mathbf{5}},\mathbf{6}\right) _{2}\oplus\left(
\overline{\mathbf{5}},\overline{\mathbf{4}}\right) _{-3}  \nonumber \\
&=&\left( \mathbf{24},\mathbf{1,1}\right) _{0,0}\oplus\left( \mathbf{1},\mathbf{%
3,1}\right) _{0,0}\oplus\left( \mathbf{1},\mathbf{1,3}\right) _{0,0}\oplus\left(
\mathbf{1},\mathbf{2,2}\right) _{0,2}\oplus \left( \mathbf{1},\mathbf{2,2}\right)
_{0,-2} \nonumber \\
&& \oplus\left( \mathbf{1},\mathbf{1,1}\right) _{0,0} \oplus\left( \mathbf{1},\mathbf{2,1}\right) _{5,-1}\oplus\left( \mathbf{1},\mathbf{%
1,2}\right) _{5,1}\oplus\left( \mathbf{1},\mathbf{2,1}\right) _{-5,1} \oplus \left(
\mathbf{1},\mathbf{1,2}\right)_{-5,-1} \nonumber \\
&& \oplus\left( \mathbf{1},\mathbf{1,1}%
\right) _{0,0} \oplus\left( \mathbf{10},\mathbf{2,1}\right) _{1,-1}\oplus\left( \mathbf{10},\mathbf{%
1,2}\right) _{1,1}\oplus\left( \mathbf{10},\mathbf{1,1}\right) _{-4,0} \nonumber \\
&& \oplus\left(
\overline{\mathbf{10}},\mathbf{2,1}\right) _{-1,1}\oplus\left( \overline{\mathbf{%
10}},\mathbf{1,2}\right) _{-1,-1}\oplus\left( \overline{\mathbf{10}},\mathbf{1,1}%
\right) _{4,0}  \nonumber \\
&&\oplus\left( \overline{\mathbf{5}},\mathbf{2,2}\right) _{2,0}\oplus\left( \overline{%
\mathbf{5}},\mathbf{1,1}\right) _{2,2}\oplus\left( \overline{\mathbf{5}},\mathbf{%
1,1}\right) _{2,-2}\oplus\left( \overline{\mathbf{5}},\mathbf{2,1}\right)
_{-3,-1}\oplus\left( \overline{\mathbf{5}},\mathbf{1,2}\right) _{-3,1}  \nonumber
\\
&&\oplus\left( \mathbf{5},\mathbf{2,2}\right) _{-2,0}\oplus\left( \mathbf{5},\mathbf{%
1,1}\right) _{-2,2}\oplus\left( \mathbf{5},\mathbf{1,1}\right) _{-2,-2}\oplus\left(
\mathbf{5},\mathbf{2,1}\right) _{3,1}\oplus\left( \mathbf{5},\mathbf{1,2}\right)
_{3,-1}.\label{1.II.a}
\end{eqnarray}%
The same calculations were also worked out for ${\bf2}.I.b$ and ${\bf2}.II.b$, which found the same results above except with opposite weights.
Thus, it holds that%
\begin{equation}
{\bf1} = {\bf 2}.I = {\bf 2}.II|_{{\bf5}\leftrightarrow\overline{{\bf5}},{\bf10}\leftrightarrow\overline{{\bf10}}},
\end{equation}
where the subscript ${\bf5}\leftrightarrow\overline{{\bf5}},{\bf10}\leftrightarrow\overline{{\bf10}}$ refers to the representations of $\mathfrak{su}_{5}$.

Next, we consider $\mathfrak{su}_{7,2}$,
\begin{equation}
\mathbf{3}:\mathfrak{e}_{8(-24)}\rightarrow \mathfrak{su}_{7,2}\rightarrow
\mathfrak{su}_{5}\oplus\mathfrak{su}_{2,2}\oplus\mathfrak{u}_{1}\rightarrow \left\{
\begin{array}{l}
\mathfrak{su}_{5}\oplus\mathfrak{so}_{3,1,a}\oplus\mathfrak{u}_{1}\oplus\mathfrak{so}_{1,1}; \\
\\
\mathfrak{su}_{5}\oplus\mathfrak{so}_{3,1,b}\oplus\mathfrak{u}_{1}\oplus\mathfrak{so}_{1,1},%
\end{array}%
\right.
\end{equation}%
where, as above, there are two non-equivalent embeddings of $%
\mathfrak{so}_{3,1}\oplus\mathfrak{so}_{1,1}$ into $\mathfrak{su}_{2,2}$ (cf. \eqref{primmo}-\eqref{seccondo}). The branching of the $\mathbf{248}$ of $%
\mathfrak{e}_{8(-24)}$ goes as follows,%
\begin{eqnarray}
\mathbf{3}.a &:&\mathbf{248}=\mathbf{80}\oplus\mathbf{84}\oplus\overline{\mathbf{84}}
\nonumber \\
&=&\left( \mathbf{24},\mathbf{1}\right) _{0}\oplus\left( \mathbf{1},\mathbf{1}%
\right) _{0}\oplus\left( \mathbf{5},\overline{\mathbf{4}}\right) _{9}\oplus\left(
\overline{\mathbf{5}},\mathbf{4}\right) _{-9}\oplus\left( \mathbf{1},\mathbf{15}%
\right) _{0}  \nonumber \\
&&\oplus\left( \mathbf{5},\mathbf{6}\right) _{-6}\oplus\left( \mathbf{10},\mathbf{4}%
\right) _{3}\oplus\left( \overline{\mathbf{10}},\mathbf{1}\right) _{12}\oplus\left(
\mathbf{1},\overline{\mathbf{4}}\right) _{-15}  \nonumber \\
&&\oplus\left( \overline{\mathbf{5}},\mathbf{6}\right) _{6}\oplus\left( \overline{%
\mathbf{10}},\overline{\mathbf{4}}\right) _{-3}\oplus\left( \mathbf{10},\mathbf{1}%
\right) _{-12}\oplus\left( \mathbf{1},\mathbf{4}\right) _{15}  \nonumber \\
&=&\left( \mathbf{24},\mathbf{1,1}\right) _{0,0}\oplus\left( \mathbf{1},\mathbf{%
3,1}\right) _{0,0}\oplus\left( \mathbf{1},\mathbf{1,3}\right) _{0,0}\oplus\left(
\mathbf{1},\mathbf{2,2}\right) _{0,2}\oplus\left( \mathbf{1},\mathbf{2,2}\right)
_{0,-2}  \nonumber \\
&& \oplus\left( \mathbf{1},\mathbf{1,1}\right) _{0,0} \oplus\left( \mathbf{1},\mathbf{2,1}\right) _{15,1}\oplus\left( \mathbf{1},\mathbf{%
1,2}\right) _{15,-1}\oplus\left( \mathbf{1},\mathbf{2,1}\right) _{-15,-1}\oplus\left(
\mathbf{1},\mathbf{1,2}\right) _{-15,1} \nonumber \\
&& \oplus\left( \mathbf{1},\mathbf{1,1}%
\right) _{0,0} \oplus\left( \overline{\mathbf{10}},\mathbf{2,1}\right) _{-3,-1}\oplus\left(
\overline{\mathbf{10}},\mathbf{1,2}\right) _{-3,1}\oplus\left( \overline{\mathbf{%
10}},\mathbf{1,1}\right) _{12,0} \nonumber \\
&& \oplus\left( \mathbf{10},\mathbf{2,1}\right)
_{3,1}\oplus\left( \mathbf{10},\mathbf{1,2}\right) _{3,-1}\oplus\left( \mathbf{10},%
\mathbf{1,1}\right) _{-12,0}  \nonumber \\
&&\oplus\left( \overline{\mathbf{5}},\mathbf{2,2}\right) _{6,0}\oplus\left( \overline{%
\mathbf{5}},\mathbf{1,1}\right) _{6,2}\oplus\left( \overline{\mathbf{5}},\mathbf{%
1,1}\right) _{6,-2}\oplus\left( \overline{\mathbf{5}},\mathbf{2,1}\right)
_{-9,1}\oplus\left( \overline{\mathbf{5}},\mathbf{1,2}\right) _{-9,-1}  \nonumber
\\
&&\oplus\left( \mathbf{5},\mathbf{2,2}\right) _{-6,0}\oplus\left( \mathbf{5},\mathbf{%
1,1}\right) _{-6,2}\oplus\left( \mathbf{5},\mathbf{1,1}\right) _{-6,-2}\oplus\left(
\mathbf{5},\mathbf{2,1}\right) _{9,-1}\oplus\left( \mathbf{5},\mathbf{1,2}\right)
_{9,1}.\label{2.a}
\end{eqnarray}%
\begin{eqnarray}
\mathbf{3}.b &:&\mathbf{248}=\mathbf{80}\oplus\mathbf{84}\oplus\overline{\mathbf{84}}
\nonumber \\
&=&\left( \mathbf{24},\mathbf{1}\right) _{0}\oplus\left( \mathbf{1},\mathbf{1}%
\right) _{0}\oplus\left( \mathbf{5},\overline{\mathbf{4}}\right) _{9}\oplus\left(
\overline{\mathbf{5}},\mathbf{4}\right) _{-9}\oplus\left( \mathbf{1},\mathbf{15}%
\right) _{0}  \nonumber \\
&&\oplus\left( \mathbf{5},\mathbf{6}\right) _{-6}\oplus\left( \mathbf{10},\mathbf{4}%
\right) _{3}\oplus\left( \overline{\mathbf{10}},\mathbf{1}\right) _{12}\oplus\left(
\mathbf{1},\overline{\mathbf{4}}\right) _{-15}  \nonumber \\
&&\oplus\left( \overline{\mathbf{5}},\mathbf{6}\right) _{6}\oplus\left( \overline{%
\mathbf{10}},\overline{\mathbf{4}}\right) _{-3}\oplus\left( \mathbf{10},\mathbf{1}%
\right) _{-12}\oplus\left( \mathbf{1},\mathbf{4}\right) _{15}  \nonumber \\
&=&\left( \mathbf{24},\mathbf{1,1}\right) _{0,0}\oplus\left( \mathbf{1},\mathbf{%
3,1}\right) _{0,0}\oplus\left( \mathbf{1},\mathbf{1,3}\right) _{0,0}\oplus\left(
\mathbf{1},\mathbf{2,2}\right) _{0,2}\oplus\left( \mathbf{1},\mathbf{2,2}\right)
_{0,-2} \nonumber \\
&& \oplus\left( \mathbf{1},\mathbf{1,1}\right) _{0,0} \oplus\left( \mathbf{1},\mathbf{2,1}\right) _{15,-1}\oplus\left( \mathbf{1},\mathbf{%
1,2}\right) _{15,1}\oplus\left( \mathbf{1},\mathbf{2,1}\right) _{-15,1}\oplus\left(
\mathbf{1},\mathbf{1,2}\right) _{-15,-1} \nonumber \\
&& \oplus\left( \mathbf{1},\mathbf{1,1}%
\right) _{0,0} \oplus\left( \mathbf{10},\mathbf{2,1}\right) _{3,-1}\oplus\left( \mathbf{10},\mathbf{%
1,2}\right) _{3,1}\oplus\left( \mathbf{10},\mathbf{1,1}\right) _{-12,0} \nonumber \\
&& \oplus\left(
\overline{\mathbf{10}},\mathbf{2,1}\right)_{-3,1}\oplus\left( \overline{\mathbf{%
10}},\mathbf{1,2}\right) _{-3,-1}\oplus\left( \overline{\mathbf{10}},\mathbf{1,1}%
\right) _{12,0}  \nonumber \\
&&\oplus\left( \overline{\mathbf{5}},\mathbf{2,2}\right) _{6,0}\oplus\left( \overline{%
\mathbf{5}},\mathbf{1,1}\right) _{6,2}\oplus\left( \overline{\mathbf{5}},\mathbf{%
1,1}\right) _{6,-2}\oplus\left( \overline{\mathbf{5}},\mathbf{2,1}\right)
_{-9,-1}\oplus\left( \overline{\mathbf{5}},\mathbf{1,2}\right) _{-9,1}  \nonumber
\\
&&\oplus\left( \mathbf{5},\mathbf{2,2}\right) _{-6,0}\oplus\left( \mathbf{5},\mathbf{%
1,1}\right) _{-6,2}\oplus\left( \mathbf{5},\mathbf{1,1}\right) _{-6,-2}\oplus\left(
\mathbf{5},\mathbf{2,1}\right) _{9,1}\oplus\left( \mathbf{5},\mathbf{1,2}\right)
_{9,-1}.\label{2.b}
\end{eqnarray}%
Thus, it holds%
\begin{eqnarray}
\left. \mathbf{3}.a\right\vert _{Q_{u_{1}}\rightarrow Q_{u_{1}}/3} &=&\left.
\mathbf{2}.I.b\right\vert _{\mathbf{5}\leftrightarrow \overline{\mathbf{5}},%
\mathbf{10}\leftrightarrow \overline{\mathbf{10}}}=\mathbf{2}.II.b, \\
\left. \mathbf{3}.b\right\vert _{Q_{u_{1}}\rightarrow Q_{u_{1}}/3} &=&\left.
\mathbf{2}.I.a\right\vert _{\mathbf{5}\leftrightarrow \overline{\mathbf{5}},%
\mathbf{10}\leftrightarrow \overline{\mathbf{10}}}=\mathbf{2}.II.a,
\end{eqnarray}%
where \textquotedblleft $Q_{u_{1}}\rightarrow Q_{u_{1}}/3$" denotes a
rescaling of the charge of $u_{1}$ by a factor $1/3$.\bigskip

\textbf{To recap:}
As shown above, $\mathfrak{so}_{10}\oplus\mathfrak{so}_{2,4}$, $\mathfrak{su}_{2,7}$, and $\mathfrak{su}_5\oplus\mathfrak{su}_{2,3}$ all can lead to the same $\mathfrak{su}_{5}\oplus \mathfrak{so}_{4,2}\oplus \mathfrak{u}_1$, which can further be broken to $\mathfrak{su}_{5}\oplus\mathfrak{so}_{3,1}\oplus\mathfrak{u}_1\oplus\mathfrak{so}_{1,1}$. Up to a rescaling of the $\mathfrak{u}_1$ charges, all of the representations from these three paths coincide with the same result.

\section{High energy theories from four timelike dimensions}
\label{4temporalDim}

Given the reluctancy to study additional timelike dimensions, there may be additional reluctancy to pursue twelve timelike dimensions, as we have considered in Sec. \eqref{Sec3}. Therefore, in this section, we will consider the maximal subalgebra $\mathfrak{so}_{12,4}$ of $\mathfrak{e}_{8(-24)}$ to have its 16-dimensional vector representation ${\bf16}$ with signature $(s,t)=(12,4)$.
In particular, we explore two possibilities to work with $\mathfrak{e}_{8(-24)}$ with four times. First, in Sec. \eqref{Sec41} we attempt to connect to graviweak unification with $\mathfrak{so}_{3,1}$ spacetime, which demonstrates some internal consistency but most likely violates the Coleman-Mandula theorem, since we work with a real form of $\mathfrak{e}_8$. Secondly, in Sec. \eqref{Sec42} we explore $\mathfrak{so}_{2,2}$ spacetime, which can be utilized as global isometry for $AdS_3$ and have no issues with the Coleman-Mandula theorem. We will primarily focus on $\mathfrak{so}_{10}$ and Pati-Salam GUTs in this section, as this allows for the easiest comparison with $\mathfrak{so}_{3,3}\oplus\mathfrak{so}_{9,1}$.

\subsection{An attempt for $SO(3,1)$}\label{Sec41}
When working with $\mathfrak{so}_{12,4}$, it is tempting to isolate $\mathfrak{so}_{3,1}$ for spacetime. This, however, leaves behind $\mathfrak{so}_{9,3}$, which does not allow for $\mathfrak{so}_{10}$ or Pati-Salam GUTs. While it could be conceivable that $\mathfrak{so}_{10}$ could overlap with $\mathfrak{so}_{3,1}$ maximally by $\mathfrak{so}_3$ or $\mathfrak{so}_2$, this would appear to violate the Coleman-Mandula theorem, as this implies that the spacetime and internal symmetries would not be a direct product. However, studying gravity as a gauge theory has allowed for clever ways to get around the Coleman-Mandula theorem~\cite{1982Hestenes,Chisholm:1986zf,Chisholm1992,Trayling:2001kd,Nesti:2007ka,HESTENES_2008}. While it is impossible to break $\mathfrak{so}_{12,4}$ to $\mathfrak{so}_{10}\oplus\mathfrak{so}_{3,1}$ with $\mathfrak{so}_{10}$ as spatial dimensions, we compare two paths of symmetry breaking that go through $\mathfrak{so}_{2,4}\oplus\mathfrak{so}_{10}$ and $\mathfrak{so}_{3,3}\oplus\mathfrak{so}_{9,1}$ and look to see if there is at least self-consistency with respect to chirality of the $\mathfrak{so}_{10}$ spinors. Since graviweak unification works with complex $\mathfrak{so}_{3,1}$~\cite{Nesti:2007ka}, it's clear that we can't recover the full graviweak unification with a real form of $\mathfrak{e}_{8}$. However, we can take two different paths of symmetry breaking and show how $\mathfrak{so}_{3,1}$ spacetime and $\mathfrak{su}_{2,L}\oplus\mathfrak{su}_{2,R}$ of Pati-Salam may overlap.

A double gauge theory that acts on spinors from the left and right to allow for gravity on one side and the electroweak symmetry on the other can be used in a Clifford/geometric algebra formalism,~\cite{1982Hestenes,Trayling:2001kd,HESTENES_2008}
\begin{equation}
\psi \rightarrow \psi' = L\psi U, \quad L = e^{\frac{1}{2}B},\quad U = e^{i\sigma_z \chi},
\end{equation}
where $L$ generates Lorentz transformations in terms of a bivector generator $B$, $\sigma_z$ is an $SU(2)$ Pauli matrix of weak isospace, and $\chi$ is a $U(1)$ hypercharge gauge parameter such that $A^\mu{}' = A^\mu - \partial^\mu \chi$. This notion of two gauge theories acting from different sides bypasses the assumptions of Coleman-Mandula~\cite{Trayling:2001kd}. While it is unclear if the two paths of symmetry breaking below are related precisely to graviweak unification or the Clifford/geometric algebra approaches, it is plausible that something similar allows for the Coleman-Mandula theorem to not be violated.

The two paths explored are
\begin{equation}
\label{two-paths}
\begin{array}{rcl}
& \mathfrak{e}_{8(-24)}& \\
& \downarrow & \\
& \mathfrak{so}_{12,4} & \\
& \swarrow\qquad \qquad \searrow & \\
\mathfrak{so}_{2,4}\oplus \mathfrak{so}_{10} &  & \mathfrak{so}_{3,3}\oplus\mathfrak{so}_{9,1} \\
\downarrow \qquad \,\,&  & \,\,\qquad \downarrow \\
\mathfrak{so}_{2,4} \oplus \mathfrak{so}_{6}\oplus\mathfrak{su}_{2;L}\oplus\mathfrak{su}_{2;R} &  & \mathfrak{so}_{3,3}\oplus\mathfrak{so}_8\oplus\mathfrak{so}_{1,1;c} \\
\downarrow \qquad \,\,&  & \,\,\qquad \downarrow \\
\mathfrak{so}_{2,4} \oplus \mathfrak{so}_6 \oplus \mathfrak{su}_{2;L}\oplus \mathfrak{u}_{1;R} & & \mathfrak{so}_{3,3}\oplus\mathfrak{so}_6\oplus\mathfrak{so}_{1,1;c}\oplus\mathfrak{so}_{2,0;c} \\
\downarrow \qquad\,\, &  & \,\,\qquad \downarrow \\
\mathfrak{so}_{2,4}\oplus\mathfrak{so}_6\oplus\mathfrak{u}_{1;L} \oplus\mathfrak{u}_{1;R} & & \mathfrak{so}_{1,3}\oplus\mathfrak{so}_6\oplus\mathfrak{so}_{1,1;c} \oplus \mathfrak{so}_{2,0;c}\oplus\mathfrak{so}_{2,0;s} \\
\downarrow\qquad\,\,  &  & \\
\mathfrak{so}_{1,3}\oplus\mathfrak{so}_6\oplus \mathfrak{so}_{1,1;c'}\oplus\mathfrak{u}_{1;L} \oplus\mathfrak{u}_{1;R} & &
\end{array}
\end{equation}
To clarify, we break $\mathfrak{su}_{2;L}\oplus\mathfrak{su}_{2;R}$ to $\mathfrak{u}_{1;L}\oplus\mathfrak{u}_{1;R}$ to compare with $\mathfrak{so}_{2,0;c}\oplus\mathfrak{so}_{2,0;s}$, where the subscripts $c$ and $s$ refer to charge space and spacetime. We also establish that $\mathfrak{so}_{1,1;c} = \mathfrak{so}_{1,1;c'}$, which allows for the bottom of the left chain above to be related to the bottom of the right chain above.

Starting with the left chain in Eq.~\eqref{two-paths} and omitting, here and below, intermediate breakings for brevity,
\begin{eqnarray}
\label{leftBreak}
\mathfrak{e}_{8(-24)} &\rightarrow& \mathfrak{so}_{12,4} \rightarrow \mathfrak{so}_{2,4}\oplus\mathfrak{so}_{10} \rightarrow \mathfrak{so}_{2,4}\oplus\mathfrak{su}_4\oplus\mathfrak{su}_{2;L}\oplus\mathfrak{su}_{2;R} \nonumber \\
&\rightarrow& \mathfrak{so}_{2,4} \oplus \mathfrak{su}_{4}\oplus\mathfrak{su}_{2;L}\oplus\mathfrak{u}_{1;R} \rightarrow \mathfrak{so}_{2,4}\oplus\mathfrak{su}_{4}\oplus\mathfrak{u}_{1;L}\oplus\mathfrak{u}_{1;R} \nonumber \\
&\rightarrow& \mathfrak{so}_{1,3}\oplus\mathfrak{su}_4\oplus\mathfrak{so}_{1,1;c'}\oplus\mathfrak{u}_{1;L}\oplus\mathfrak{u}_{1;R},  \\
{\bf248} &=& {\bf120} \oplus {\bf128} = ({\bf15},{\bf1}) \oplus ({\bf1},{\bf45}) \oplus ({\bf6},{\bf10}) \oplus ({\bf4},{\bf16}) \oplus (\overline{{\bf4}},\overline{{\bf16}}) \nonumber \\
&=& ({\bf15},{\bf1},{\bf1},{\bf1}) \oplus ({\bf1},{\bf15},{\bf1},{\bf1}) \oplus ({\bf1},{\bf1},{\bf3},{\bf1}) \oplus ({\bf1},{\bf1},{\bf1},{\bf3}) \oplus ({\bf6},{\bf6},{\bf1},{\bf1}) \oplus ({\bf6},{\bf1},{\bf2},{\bf2})  \nonumber \\
&& \oplus ({\bf1},{\bf6},{\bf2},{\bf2}) \oplus ({\bf4},{\bf4},{\bf2},{\bf1})^{L} \oplus ({\bf4},\overline{{\bf4}},{\bf1},{\bf2})^{\overline{L}} \oplus (\overline{{\bf4}},\overline{{\bf4}},{\bf2},{\bf1})^{\overline{R}} \oplus (\overline{{\bf4}},{\bf4},{\bf1},{\bf2})^R \nonumber \\
&=& ({\bf3},{\bf1},{\bf1})_{0,0,0} \oplus ({\bf1},{\bf3},{\bf1})_{0,0,0} \oplus ({\bf1},{\bf1},{\bf1})_{0,0,0} \oplus ({\bf2},{\bf2},{\bf1})_{2,0,0} \oplus ({\bf2},{\bf2},{\bf1})_{-2,0,0} \nonumber \\
&& \oplus ({\bf1},{\bf1},{\bf15})_{0,0,0} \oplus ({\bf1},{\bf1},{\bf1})_{0,0,0} \oplus ({\bf1},{\bf1},{\bf1})_{0,2,0} \oplus ({\bf1},{\bf1},{\bf1})_{0,-2,0}  \nonumber \\
&&  \oplus ({\bf1},{\bf1},{\bf1})_{0,0,0} \oplus ({\bf1},{\bf1},{\bf1})_{0,0,2} \oplus ({\bf1},{\bf1},{\bf1})_{0,0,-2} \oplus ({\bf1},{\bf1},{\bf6})_{0,1,1} \nonumber \\
&& \oplus ({\bf1},{\bf1},{\bf6})_{0,-1,1} \oplus ({\bf1},{\bf1},{\bf6})_{0,1,-1} \oplus ({\bf1},{\bf1},{\bf6})_{0,-1,-1} \oplus ({\bf2},{\bf2},{\bf6})_{0,0,0} \nonumber \\
&& \oplus ({\bf1},{\bf1},{\bf6})_{2,0,0} \oplus ({\bf1},{\bf1},{\bf6})_{-2,0,0} \oplus ({\bf2},{\bf2},{\bf1})_{0,1,1} \oplus ({\bf1},{\bf1},{\bf1})_{2,1,1}  \nonumber \\
&&\oplus ({\bf1},{\bf1},{\bf1})_{-2,1,1} \oplus ({\bf2},{\bf2},{\bf1})_{0,-1,1} \oplus ({\bf1},{\bf1},{\bf1})_{2,-1,1} \oplus ({\bf1},{\bf1},{\bf1})_{-2,-1,1} \nonumber \\
&& \oplus ({\bf2},{\bf2},{\bf1})_{0,1,-1} \oplus ({\bf1},{\bf1},{\bf1})_{2,1,-1}  \oplus ({\bf1},{\bf1},{\bf1})_{-2,1,-1} \nonumber \\
&& \oplus ({\bf2},{\bf2},{\bf1})_{0,-1,-1} \oplus ({\bf1},{\bf1},{\bf1})_{2,-1,-1} \oplus ({\bf1},{\bf1},{\bf1})_{-2,-1,-1} \oplus ({\bf2},{\bf1},{\bf4})_{1,1,0}^{L} \nonumber \\
&& \oplus ({\bf1},{\bf2},{\bf4})_{-1,1,0}^{LM} \oplus ({\bf2},{\bf1},{\bf4})_{1,-1,0}^{L} \oplus ({\bf1},{\bf2},{\bf4})_{-1,-1,0}^{LM} \oplus ({\bf2},{\bf1},\overline{{\bf4}})_{1,0,1}^{\overline{L}} \nonumber \\
&& \oplus ({\bf1},{\bf2},\overline{{\bf4}})_{-1,0,1}^{\overline{L}M} \oplus ({\bf2},{\bf1},\overline{{\bf4}})_{1,0,-1}^{\overline{L}} \oplus ({\bf1},{\bf2},\overline{{\bf4}})_{-1,0,-1}^{\overline{L}M}  \nonumber \\
&& \oplus ({\bf1},{\bf2},\overline{{\bf4}})_{1,1,0}^{\overline{R}} \oplus ({\bf2},{\bf1},\overline{{\bf4}})_{-1,1,0}^{\overline{R}M} \oplus ({\bf1},{\bf2},\overline{{\bf4}})_{1,-1,0}^{\overline{R}} \oplus ({\bf2},{\bf1},\overline{{\bf4}})_{-1,-1,0}^{\overline{R}M}  \nonumber \\
&& \oplus ({\bf1},{\bf2},{\bf4})_{1,0,1}^{R} \oplus ({\bf2},{\bf1},{\bf4})_{-1,0,1}^{RM} \oplus ({\bf1},{\bf2},{\bf4})_{1,0,-1}^{R} \oplus ({\bf2},{\bf1},{\bf4})_{-1,0,-1}^{RM}, \nonumber
\end{eqnarray}
where the superscripts $L,\overline{L},R,$ and $\overline{R}$ denote $\psi_L$, $\overline{\psi}_L$, $\overline{\psi}_R$, and $\psi_R$ of Pati-Salam, respectively, and $M$ corresponds to mirror fermions.

Next, we focus on the right chain in Eq.~\eqref{two-paths},
\begin{eqnarray}
\label{rightBreak}
\mathfrak{e}_{8(-24)} &\rightarrow& \mathfrak{so}_{12,4} \rightarrow \mathfrak{so}_{3,3}\oplus\mathfrak{so}_{9,1} \rightarrow \mathfrak{so}_{3,3}\oplus\mathfrak{so}_8\oplus\mathfrak{so}_{1,1;c} \rightarrow \mathfrak{so}_{3,3}\oplus\mathfrak{so}_6\oplus\mathfrak{so}_{1,1;c}\oplus\mathfrak{so}_{2,0;c} \nonumber \\
&\rightarrow& \mathfrak{so}_{1,3}\oplus\mathfrak{so}_6\oplus\mathfrak{so}_{1,1;c}\oplus\mathfrak{so}_{2,0;c}\oplus\mathfrak{so}_{2,0;s}, \\
{\bf248} &=& {\bf120} \oplus {\bf128} = ({\bf15},{\bf1}) \oplus ({\bf1},{\bf45}) \oplus ({\bf6},{\bf10}) \oplus ({\bf4},{\bf16}) \oplus ({\bf4}',{\bf16}') \nonumber \\
&=& ({\bf3},{\bf1},{\bf1})_{0,0,0} \oplus ({\bf1},{\bf3},{\bf1})_{0,0,0} \oplus ({\bf1},{\bf1},{\bf1})_{0,0,0} \nonumber \\
&& \oplus ({\bf2},{\bf2},{\bf1})_{0,0,2} \oplus ({\bf2},{\bf2},{\bf1})_{0,0,-2} \oplus ({\bf1},{\bf1},{\bf15})_{0,0,0} \oplus ({\bf1},{\bf1},{\bf1})_{0,0,0} \nonumber\\
&& \oplus ({\bf1},{\bf1},{\bf6})_{0,2,0} \oplus ({\bf1},{\bf1},{\bf6})_{0,-2,0} \oplus ({\bf1},{\bf1},{\bf1})_{0,0,0} \oplus ({\bf1},{\bf1},{\bf6})_{2,0,0} \nonumber \\
&& \oplus ({\bf1},{\bf1},{\bf1})_{2,2,0} \oplus ({\bf1},{\bf1},{\bf1})_{2,-2,0} \oplus ({\bf1},{\bf1},{\bf6})_{-2,0,0} \oplus ({\bf1},{\bf1},{\bf1})_{-2,2,0} \nonumber \\
&&  \oplus ({\bf1},{\bf1},{\bf1})_{-2,-2,0} \oplus ({\bf2},{\bf2},{\bf6})_{0,0,0} \oplus ({\bf1},{\bf1},{\bf6})_{0,0,2} \oplus ({\bf1},{\bf1},{\bf6})_{0,0,-2} \nonumber \\
&& \oplus ({\bf2},{\bf2},{\bf1})_{0,2,0} \oplus ({\bf1},{\bf1},{\bf1})_{0,2,2} \oplus ({\bf1},{\bf1},{\bf1})_{0,2,-2} \oplus ({\bf2},{\bf2},{\bf1})_{0,-2,0} \nonumber \\
&& \oplus ({\bf1},{\bf1},{\bf1})_{0,-2,2} \oplus ({\bf1},{\bf1},{\bf1})_{0,-2,-2} \oplus ({\bf2},{\bf2},{\bf1})_{2,0,0} \oplus ({\bf1},{\bf1},{\bf1})_{2,0,2} \nonumber \\
&& \oplus ({\bf1},{\bf1},{\bf1})_{2,0,-2} \oplus ({\bf2},{\bf2},{\bf1})_{-2,0,0} \oplus ({\bf1},{\bf1},{\bf1})_{-2,0,2} \oplus ({\bf1},{\bf1},{\bf1})_{-2,0,-2} \nonumber \\
&& \oplus ({\bf2},{\bf1},\overline{{\bf4}})_{1,1,1} \oplus ({\bf1},{\bf2},\overline{{\bf4}})_{1,1,-1} \oplus ({\bf2},{\bf1},{\bf4})_{1,-1,1} \oplus ({\bf1},{\bf2},{\bf4})_{1,-1,-1} \nonumber \\
&& \oplus ({\bf2},{\bf1},{\bf4})_{-1,1,1} \oplus ({\bf1},{\bf2},{\bf4})_{-1,1,-1} \oplus ({\bf2},{\bf1},\overline{{\bf4}})_{-1,-1,1} \oplus ({\bf1},{\bf2},\overline{{\bf4}})_{-1,-1,-1} \nonumber \\
&& \oplus ({\bf1},{\bf2},{\bf4})_{1,1,1} \oplus ({\bf2},{\bf1},{\bf4})_{1,1,-1} \oplus ({\bf1},{\bf2},\overline{{\bf4}})_{1,-1,1} \oplus ({\bf2},{\bf1},\overline{{\bf4}})_{1,-1,-1} \nonumber \\
&& \oplus ({\bf1},{\bf2},\overline{{\bf4}})_{-1,1,1} \oplus ({\bf2},{\bf1},\overline{{\bf4}})_{-1,1,-1} \oplus ({\bf1},{\bf2},{\bf4})_{-1,-1,1} \oplus ({\bf2},{\bf1},{\bf4})_{-1,-1,-1} \nonumber
\end{eqnarray}

Comparing Eqs.~\eqref{leftBreak} and \eqref{rightBreak}, the representations are identical, besides the charges. However, they can be related to each other by
\begin{equation}
Q_L = \frac{1}{2}\left(Q_c - Q_s \right), \qquad Q_R = \frac{1}{2}\left(Q_c + Q_s \right),
\end{equation}
where $Q_L$ and $Q_R$ are the charges from Eq.~\eqref{leftBreak}, while $Q_c$ and $Q_s$ are the charges from Eq.~\eqref{rightBreak}. Furthermore, the weights associated with $\mathfrak{so}_{1,1;c}$ and $\mathfrak{so}_{1,1;c'}$ are identical. This demonstrates that the two paths from Eq.~\eqref{two-paths} are identical up to rescaling of the charges shown above.

Next, we would like to ensure that the fermionic spinors have a self-consistent chirality with respect to $\mathfrak{so}_{10}$ and $\mathfrak{so}_{3,1}$. Since it was impossible to isolate $\mathfrak{so}_{10}\oplus\mathfrak{so}_{3,1}$, the right path in Eq.~\eqref{rightBreak} broke $\mathfrak{so}_{3,3}$ spacetime to $\mathfrak{so}_{1,3}\oplus\mathfrak{so}_{2,0;s}$, which allowed for us to ensure that the representations matched with those found in Eq.~\eqref{leftBreak}. This $\mathfrak{so}_{1,3}$ has one spacelike and three timelike dimensions, and so it is not spacetime; however, we look to break $\mathfrak{so}_{3,3}$ in two paths to understand the appropriate chiralities with respect to $\mathfrak{so}_{3,1}$,
\begin{equation}
\begin{array}{ccc}
& (A): \quad \mathfrak{so}_{3,1}\oplus\mathfrak{so}_{1,1;c}\oplus\mathfrak{so}_{0,2} & \\
\mathfrak{so}_{3,3}\oplus\mathfrak{so}_{1,1;c}\quad{}^\nearrow_\searrow & &{}^\searrow_\nearrow\quad \mathfrak{so}_{1,1;c}\oplus\mathfrak{so}_{2,0;s}\oplus\mathfrak{so}_{0,2}\oplus\mathfrak{so}_{1,1;s} \\
& (B):\, \mathfrak{so}_{1,3}\oplus\mathfrak{so}_{1,1;c}\oplus\mathfrak{so}_{2,0;s}&
\end{array}
\end{equation}
The $\mathfrak{so}_{1,1;c}$ factor is included to help keep track of mirror fermions. Disregarding $\mathfrak{so}_8$ and its associated reps, we focus on a subclass of fermions that corresponds to the ones studied in Section~\eqref{3massGen}, giving
\begin{equation}
32 = {\bf15}_0 \oplus {\bf1}_0 \oplus {\bf4}_1 \oplus {\bf4}_{-1}^M \oplus {\bf4}'_1 \oplus {\bf4}'{}^M_{-1}
\end{equation}
where these representations are for $\mathfrak{so}_{3,3}\oplus\mathfrak{so}_{1,1;c}$.
Note that $32$ is not a formal representation for any algebra, but we use it as a compact way to refer to these 32 dofs.

Focusing on the top path $(A)$,
\begin{eqnarray}
(A): && \mathfrak{so}_{3,1}\oplus\mathfrak{so}_{1,1;c}\oplus\mathfrak{so}_{0,2} \rightarrow \mathfrak{so}_{1,1;c} \oplus \mathfrak{so}_{2,0;s}\oplus\mathfrak{so}_{0,2}\oplus\mathfrak{so}_{1,1;s}, \\
32 &=& ({\bf 3},{\bf1})_{0,0} \oplus ({\bf1},{\bf3})_{0,0} \oplus ({\bf1},{\bf1})_{0,0} \oplus ({\bf2},{\bf2})_{0,2}\oplus({\bf2},{\bf2})_{0,-2} \oplus ({\bf 1},{\bf1})_{0,0} \nonumber \\
&& \oplus ({\bf2},{\bf1})_{1,1}^L \oplus ({\bf1},{\bf2})_{1,-1}^R \oplus ({\bf2},{\bf1})_{-1,1}^{LM}\oplus ({\bf1},{\bf2})_{-1,-1}^{RM} \nonumber \\
&& \oplus ({\bf1},{\bf2})_{1,1}^R \oplus ({\bf2},{\bf1})_{1,-1}^L \oplus ({\bf1},{\bf2})_{-1,1}^{RM} \oplus ({\bf2},{\bf1})_{-1,-1}^{LM} \nonumber \\
&=& {\bf1}_{0,0,0,0}\oplus {\bf1}_{0,2,0,0}\oplus {\bf1}_{0,-2,0,0}\oplus {\bf1}_{0,0,0,0}\oplus {\bf1}_{0,0,0,2}\oplus {\bf1}_{0,0,0,-2} \nonumber \\
&& \oplus {\bf1}_{0,0,0,0}\oplus {\bf1}_{0,0,0,0} \oplus {\bf1}_{0,2,1,1}\oplus {\bf1}_{0,2,1,-1}\oplus {\bf1}_{0,-2,1,1}\oplus {\bf1}_{0,-2,1,-1}\oplus {\bf1}_{0,2,-1,1} \nonumber \\
&& \oplus {\bf1}_{0,2,-1,-1}\oplus {\bf1}_{0,-2,-1,1}\oplus {\bf1}_{0,-2,-1,-1} \oplus {\bf1}_{1,1,1,0}^L\oplus {\bf1}_{1,-1,1,0}^L\oplus {\bf1}_{1,0,-1,1}^R  \nonumber \\
&& \oplus {\bf1}_{1,0,-1,-1}^R\oplus {\bf1}_{-1,1,1,0}^{LM}\oplus {\bf1}_{-1,-1,1,0}^{LM}\oplus {\bf1}_{-1,0,-1,1}^{RM}\oplus {\bf1}_{-1,0,-1,-1}^{RM} \oplus {\bf1}_{1,0,1,1}^R  \nonumber \\
&& \oplus {\bf1}_{1,0,1,-1}^R\oplus {\bf1}_{1,1,-1,0}^L\oplus {\bf1}_{1,-1,-1,0}^L\oplus {\bf1}_{-1,0,1,1}^{RM}\oplus {\bf1}_{-1,0,1,-1}^{RM}\oplus {\bf1}_{-1,1,-1,0}^{LM}\oplus {\bf1}_{-1,-1,-1,0}^{LM}. \nonumber
\end{eqnarray}
This allows us to ensure which fermionic dofs are associated with left and right chiralities, which are labelled by the superscripts $L$ and $R$.

Focusing on the bottom path $(B)$,
\begin{eqnarray}
(B): && \mathfrak{so}_{1,3}\oplus\mathfrak{so}_{1,1;c}\oplus\mathfrak{so}_{2,0;s} \rightarrow \mathfrak{so}_{1,1;c}\oplus\mathfrak{so}_{2,0;s}\oplus\mathfrak{so}_{0,2}\oplus\mathfrak{so}_{1,1;s}, \\
32 &=& ({\bf 3},{\bf1})_{0,0} \oplus ({\bf1},{\bf3})_{0,0} \oplus ({\bf1},{\bf1})_{0,0} \oplus ({\bf2},{\bf2})_{0,2}\oplus({\bf2},{\bf2})_{0,-2} \oplus ({\bf 1},{\bf1})_{0,0} \nonumber \\
&& \oplus ({\bf2},{\bf1})_{1,1}^L \oplus ({\bf1},{\bf2})_{1,-1}^R \oplus ({\bf2},{\bf1})_{-1,1}^{LM}\oplus ({\bf1},{\bf2})_{-1,-1}^{RM} \nonumber \\
&& \oplus ({\bf1},{\bf2})_{1,1}^R \oplus ({\bf2},{\bf1})_{1,-1}^L \oplus ({\bf1},{\bf2})_{-1,1}^{RM} \oplus ({\bf2},{\bf1})_{-1,-1}^{LM} \nonumber \\
&=& {\bf1}_{0,0,0,0}\oplus {\bf1}_{0,0,2,0}\oplus {\bf1}_{0,0,-2,0}\oplus {\bf1}_{0,0,0,0}\oplus {\bf1}_{0,0,0,2} \oplus {\bf1}_{0,0,0,-2} \nonumber \\
&& \oplus {\bf1}_{0,0,0,0}\oplus {\bf1}_{0,0,0,0} \oplus {\bf1}_{0,2,1,1}\oplus {\bf1}_{0,2,1,-1}\oplus {\bf1}_{0,2,-1,1} \oplus {\bf1}_{0,2,-1,-1} \oplus {\bf1}_{0,-2,1,1} \nonumber \\
&& \oplus {\bf1}_{0,-2,1,-1}\oplus {\bf1}_{0,-2,-1,1}\oplus {\bf1}_{0,-2,-1,-1} \oplus {\bf1}_{1,1,1,0}^L\oplus {\bf1}_{1,1,-1,0}^L \oplus {\bf1}_{1,-1,0,1}^R \nonumber \\
&& \oplus {\bf1}_{1,-1,0,-1}^R\oplus {\bf1}_{-1,1,1,0}^{LM}\oplus {\bf1}_{-1,1,-1,0}^{LM}\oplus {\bf1}_{-1,-1,0,1}^{RM}\oplus {\bf1}_{-1,-1,0,-1}^{RM} \oplus {\bf1}_{1,1,0,1}^R \nonumber \\
&& \oplus {\bf1}_{1,1,0,-1}^R\oplus {\bf1}_{1,-1,1,0}^L\oplus {\bf1}_{1,-1,-1,0}^L\oplus {\bf1}_{-1,1,0,1}^{RM}\oplus {\bf1}_{-1,1,0,-1}^{RM}\oplus {\bf1}_{-1,-1,1,0}^{LM}\oplus {\bf1}_{-1,-1,-1,0}^{LM}. \nonumber
\end{eqnarray}
Since the two paths $(A)$ and $(B)$ lead to the same representations (so long as the two $\mathfrak{u}_1$ charges are swapped), we can understand how to label the fermionic roots with $L$ and $R$ superscripts for chirality of $\mathfrak{so}_{3,1}$, even though representations of $\mathfrak{so}_{1,3}$ are found. As it turns out, $({\bf2},{\bf1})$ of $\mathfrak{so}_{1,3}$ corresponds to a left-handed chirality.

Now, we look back at Eq.~\eqref{leftBreak} and remember that the ${\bf 16}$'s of $\mathfrak{so}_{10}$ with positive $\mathfrak{so}_{1,1;c'}$ weight are left-handed, while the negative weight gives mirrors that would be right-handed. Furthermore, we look at Eq.~\eqref{rightBreak} and see that representations of $\mathfrak{so}_{1,3}$ allow for us to identify chiralities of the fermions. Comparing the representations from both paths allows us to determine that the chiralities of the fermions are self-consistent in the sense that the fermionic representations of $\mathfrak{so}_{10}$ have the desired chiralities as would be found with $\mathfrak{so}_{3,1}$.

In closing of this subsection, we do not make any claims whether $\mathfrak{e}_{8(-24)}$ can be utilized in this way to give a realistic model that bypasses the Coleman-Mandula theorem, but figured it was worthwhile to at least demonstrate similarities with graviweak unification~\cite{Nesti:2007ka,Alexander:2007mt,Percacci:2008zf,Das:2013xha,Froggatt:2013lba,Das:2014xga,Laperashvili:2014xea,Laperashvili:2015pea,Das:2015usa,Sidharth:2018zev}. Graviweak unification does allow for a nontrivial way to have the weak force overlapping with spacetime without violating the Coleman-Mandula theorem, so perhaps this work will be inspirational for future work to address if $\mathfrak{e}_{8(-24)}$ can be used in this manner.


%

\subsection{Spacetime from $AdS_3$}\label{Sec42}

In order to refer to the gauge symmetry via Pati-Salam or $\mathfrak{so}_{10}$ GUT, $\mathfrak{so}_{12,4}$ must be broken to $\mathfrak{so}_{2,4}$, which can be regarded as the global isometry of a timelike $AdS_5$ for three generations, or to $\mathfrak{so}_{2,2}$ spacetime for a single generation with $\mathfrak{so}_{10,2}$ charge algebra internally, which can give an $AdS_3\times AdS_{11}$ dual symmetry. Alternatively, $SO(10,2)$ can be thought of as the conformal symmetry of $SO(9,1)$.

Breaking from $\mathfrak{e}_{8(-24)}$ to $\mathfrak{so}_{2,2}\oplus\mathfrak{so}_{10}\oplus\mathfrak{u}_1$ gives
\begin{eqnarray}
\mathfrak{e}_{8(-24)} &\rightarrow& \mathfrak{so}_{12,4} \rightarrow \mathfrak{so}_{2,2} \oplus \mathfrak{so}_{10,2} \rightarrow \mathfrak{so}_{2,2} \oplus \mathfrak{so}_{10} \oplus \mathfrak{u}_1 \nonumber \\
{\bf 248} &=& {\bf 120} \oplus {\bf 128} \nonumber \\
&=& ({\bf 3},{\bf 1},{\bf 1}) \oplus ({\bf 1},{\bf 3},{\bf 1}) \oplus ({\bf 1},{\bf 1},{\bf 66}) \oplus ({\bf 2},{\bf 2},{\bf 12}) \oplus ({\bf 2},{\bf 1},{\bf 32}) \oplus ({\bf 1},{\bf 2},{\bf 32}') \\
&=& ({\bf 3},{\bf 1},{\bf 1})_0 \oplus ({\bf 1},{\bf 3},{\bf 1})_0 \oplus ({\bf 1},{\bf 1},{\bf 45})_0 \oplus ({\bf 1},{\bf 1},{\bf 1})_0  \nonumber \\
&& \oplus ({\bf 1},{\bf 1},{\bf 10})_2 \oplus ({\bf 1},{\bf 1},{\bf 10})_{-2} \oplus ({\bf 2},{\bf 2},{\bf 10})_0 \oplus ({\bf 2},{\bf 2},{\bf 1})_2 \oplus ({\bf 2},{\bf 2},{\bf 1})_{-2} \nonumber \\
&& \oplus ({\bf 2},{\bf 1},{\bf 16})_1 \oplus ({\bf 2},{\bf 1},\overline{{\bf 16}})_{-1} \oplus ({\bf 1},{\bf 2},\overline{{\bf 16}})_1 \oplus ({\bf 1},{\bf 2},{\bf 16})_{-1}. \nonumber
\end{eqnarray}
From here, breaking to $SU(5)$ or Pati-Salam GUTs can be pursued to lead to the SM. While $AdS_5$ seems bizarre in this case, $AdS_3$ is a possibility given its relation to $D=3$ gravity with a CFT boundary theory\footnote{Note that 4D gravity can be generated from 3D gravity at high energies~\cite{Sugamoto:2001uk}. Since $\mathfrak{so}_{10}$ is a high energy GUT, considering this in 3D spacetime may allow for low-energy physics in 4D.}. 

\subsection{Branes and GUT symmetry breaking: a glance to the geometric perspective}

Within this framework, Pati-Salam GUT and the resulting SM emerging from $SO(10)$ is placed in a modern string perspective. The D3-brane in $D=9+1$ type IIB supergravity, which comes from F-theory, has a near-horizon geometry of $AdS_5\times S^5$ \cite{Chu:2016uwi}. It was shown by Sezgin, Rudychev, and Sundell that the $D=11+3$, $(1,0)$ superalgebra can reduce to the $D=9+1$ type IIA, IIB and heterotic superalgebras, as well as to the $\mathcal{N}=1$ superalgebras for $D=11+1$ and 10+1~\cite{Rudychev:1997ue,Rudychev:1997ui}. The $D=11+3$, $(1,0)$ superalgebra supports a 3-brane and 7-brane, where the 3-brane can reduce to the 3-brane of F-theory in $D=11+1$ along a single time projection.  In $D=11+1$, the 3-brane near horizon geometry is $AdS_5\times S^7$, while the 7-brane has $AdS_9\times S^3$ near horizon geometry.  We can see these geometries from breaking $SO(12,4)\rightarrow SO(4,2)\times SO(8)$ or breaking $SO(12,4)\rightarrow SO(8,2)\times SO(4)$, respectively.  Projecting to a 1-brane slice of the 3-brane, one recovers $AdS_3\times S^9$ near horizon geometry, which can be recovered from breaking $SO(12,4)\rightarrow SO(2,4)\times SO(10)$.  This projection can be done three different ways along each spatial direction of the 3-brane.

If one reduces $S^9$ of $AdS_3\times S^9$ with respect to $S^3$ of the 7-brane near horizon geometry, one has the isometry breaking $SO(10)\rightarrow SO(6)\times SO(4)$, giving the sphere decomposition $S^9 \rightarrow S^5\times S^3$.  From here, projecting $S^5\rightarrow \mathbb{CP}^2$ induces $SO(6)\sim SU(4)\rightarrow SU(3)\times U(1)_{B-L}$, while projecting $S^3\rightarrow \mathbb{CP}^1$ induces $SO(4)\rightarrow SU(2)\times U(1)$. Both of these have corresponding fibrations
\begin{eqnarray}
S^5 &\xrightarrow{S^1}& \mathbb{CP}^2, \\
S^3 &\xrightarrow{S^1}& S^2 \sim \mathbb{CP}^1 ,\nonumber
\end{eqnarray}
over $S^1$ fibers, as $SU(3)=\mbox{Isom}(\mathbb{CP}^2)$, $SU(2)=\mbox{Isom}(\mathbb{CP}^1)$, and $U(1) = \mbox{Isom}(S^1)$.  This provides a consistent geometric justification for the breaking of $\mathfrak{so}_{10}$ GUT symmetry down to the SM.

\section{Summary and Conclusions}
\label{conc}

\subsection{Summary} 
In summary, we have demonstrated that the most exceptional Lie algebra $\mathfrak{e}_8$ has precisely one non-compact real form, namely the quaternionic form $\mathfrak{e}_{8(-24)}$, that allows for the combination of spacetime Lorentz symmetry with various GUTs. We have explored both the possible signatures of the $\mathfrak{so}_{12,4}$ maximal and symmetric subalgebra of $\mathfrak{e}_{8(-24)}$: namely, the cases with twelve timelike dimensions in Sec.~\eqref{Sec3}, and the cases with four timelike dimensions in Sec. \eqref{4temporalDim}. In particular, the $\mathfrak{so}_{4,2}$ subalgebra of the maximal and symmetric subalgebra $\mathfrak{so}_{12,4}$ of $\mathfrak{e}_{8(-24)}$ could be used as a conformal symmetry or for $AdS_5$ in models with twelve timelike dimensions, while $\mathfrak{so}_{2,2}$ can be found as the isometry of $AdS_3$ in models with four timelike dimensions. Both pictures may allow for a holographic description, leading to $AdS_5/CFT_4$ and $AdS_3/CFT_2$ holography, respectively. On one hand, the models with twelve timelike dimensions may naturally reduce $AdS_5$ to $dS_4$, and thus relate to our physical universe. On the other hand, the models with four timelike dimensions may allow for a computationally tractable way to stitch together 3D gravity results to obtain 4D physics, as a 4D Riemannian manifold with local affine charts can be regarded as affine transformations of copies of $\mathbb{CP}^1$; vertex operator algebras may be useful for stitching together multiple copies of $\mathbb{CP}^1$ to obtain 4D gravity from 3D.

We obtained $SO(10)$, $SU(5)$, and Pati-Salam GUTs in both possible signatures of $\mathfrak{so}_{12,4}\subset \mathfrak{e}_{8(-24)}$, obtaining a class of generalized graviGUT models~\cite{Nesti:2009kk}, whose one has been quite recently considered in ~\cite{2014Douglas}. Moreover, we have expanded on this, by demonstrating how a Higgs scalar with three generations can be found. In Sec.~\eqref{PS} a new path of unification that bypasses $SO(10)$ and goes directly to Pati-Salam GUT was proposed, which may allow for a high energy theory that has no proton decay. In Sec.~\eqref{Sec3}, we also have proposed two new paths for $SU(5)$ GUT with spacetime, respectively starting from the maximal and non-symmetric subalgebras $\mathfrak{su}_{2,7}$ and $\mathfrak{su}_{5}\oplus\mathfrak{su}_{3,2}$ of $\mathfrak{e}_{8(-24)}$.

While $E_6$ and exceptional Jordan algebras are found in these models, these seem to differ from various approaches, such as $E_6$ GUT~\cite{Gursey1980,BARBIERI1980,Schwichtenberg:2017xhv} and recent attempts to connect $J_{3,\mathbb{O}}$ to the SM~\cite{Dubois-Violette:2016kzx,Todorov:2018mwd,Dubois-Violette:2018wgs,Todorov:2018yvi,Krasnov:2019auj,Furey:2018yyy}. Instead, we find the Peirce decomposition to give bosons and fermions, rather than only fermions\footnote{Recent work by and private communications with Dubois-Violette and Todorov~\cite{Todorov:2019hlc} suggest that the appropriate utilization of $\mathfrak{so}_{9}$ in Refs.~\cite{Dubois-Violette:2016kzx,Todorov:2018mwd,Dubois-Violette:2018wgs,Todorov:2018yvi,Krasnov:2019auj} is similar to the $\mathfrak{so}_9$ inside $\mathfrak{so}_{9,1}$ discussed in Eq.~\eqref{two-paths}.}. This is intuitive from the perspective of string theory. It still remains an open question if these class of $\mathfrak{e}_{8(-24)}$ models suggest a new $\mathfrak{e}_6$ GUT, or if the $\mathfrak{e}_6$ algebra allows for a convenient packaging of GUTs, similar to $\mathfrak{e}_{8(-24)}$.

In future work, we look to establish a Lagrangian formalism for at least one of these models. While it may appear that these models contain many additional bosonic dofs outside the SM and spacetime, this may not be the case. Note that these models account for all of the off-shell dofs of the fermions, yet the symmetry-breaking analysis of GUTs merely counts the adjoint dofs, not the bosonic off-shell dofs. It may turn out that $\mathfrak{e}_{8(-24)}$ nontrivially accounts for off-shell bosonic degrees of the SM as well, which warrants more careful study in future work.

Various phenomenological aspects such as neutrino masses and mass/flavor mixing also warrant additional study. Since the mirror electron was identified as borrowing on-shell dofs from the muon and tau, it is conceivable that $\mathfrak{e}_{8(-24)}$ may also allow for mass and flavor eigenstates.

Additionally, the exploration of charge space and its relevance for the origins of the double copy~\cite{Bern:2010ue} and KLT relations~\cite{Kawai:1985xq} is warranted, as a dual Lorentz symmetry~\cite{Cheung:2016say,Cheung:2017kzx} is found between spacetime and charge space. The notion of $\mathfrak{so}_8$ triality~\cite{Baez:2009xt} and $\mathfrak{so}_{1,9}$ charge space is suggestive of a new type of supersymmetry. If these models do allow for supersymmetry, it is clear that it is a type of charge space supersymmetry, rather than spacetime supersymmetry. This seems to differ, as additional unphysical superparticles are not needed to be introduced. This may suggest a way to break spacetime supersymmetry while preserving a charge space (i.e. internal) supersymmetry. However, it still remains unclear if these models actually contain supersymmetry or not, which should be investigated further.

Finally, the algebras occurring in exceptional periodicity (and stemming from suitable generalizations of the magic star projection) allow for a natural way to generalize $\mathfrak{e}_{8(-24)}$ that is distinct from the infinite-dimensional Kac-Moody algebras~\cite{Truini:2017jiy,Truini:2018bbd}. Given their apparent ability to describe a monstrous M-theory that adds fermions to bosonic M-theory~\cite{Rios:2019rfc}, further work is warranted to study BSM physics in relation to brane dynamics similar to those studied in generalizations of M-theory, such as F-theory and beyond~\cite{Vafa:1996xn,Bars:1996ab,Bars:1997ug,Rudychev:1997ui,Bars:2010zw,Rios:2018lhc}.

\subsection{Conclusions}

In conclusion, the Lie algebra $\mathfrak{e}_{8(-24)}$ has representation theory that has applications for model building for beyond-the-standard-model physics including gravity. Various subalgebras allow for gauge groups of the most common GUT models, including $SU(5)$, $Spin(10)$, and $SU(4)\times SU(2)\times SU(2)$. Lorentz and conformal spacetime symmetries are also found within $E_{8(-24)}$. The ${\bf128}$ Majorana-Weyl spinor representation from $E_{8(-24)}/Spin(12,4)$ allows for an efficient way to encode three generations of the standard model fermions.

\vspace{6pt}



\authorcontributions{Conceptualization, D.C, A.M. and M.R.; Methodology, D.C, A.M. and M.R.; Validation, D.C, A.M. and M.R.; Formal Analysis, D.C, A.M. and M.R.; Investigation, D.C, A.M. and M.R.; Writing – Original Draft Preparation, D.C and A.M.; Writing – Review \& Editing, D.C, A.M. and M.R.}

\funding{This research received no external funding.}

\institutionalreview{Not applicable.}

\informedconsent{Not applicable.}

\dataavailability{Not applicable.}

\acknowledgments{Thanks to Piero Truini for encouraging discussions.}

\conflictsofinterest{The authors declare no conflict of interest.}



\abbreviations{Abbreviations}{
The following abbreviations are used in this manuscript:\\

\noindent
\begin{tabular}{@{}ll}
MDPI & Multidisciplinary Digital Publishing Institute\\
DOAJ & Directory of open access journals \\
SM & standard model \\
GUT & grand unified theory \\
dofs & degrees of freedom \\
EP & exceptional periodicity
\end{tabular}
}

\appendixtitles{no} 
\appendixstart
\appendix

\begin{adjustwidth}{-\extralength}{0cm}

\reftitle{References}

\end{adjustwidth}
\end{document}